\documentclass{aa}  

\usepackage{tabularx}
\usepackage{graphicx}
\usepackage{txfonts}
\usepackage{hyperref}
\hypersetup{colorlinks=true, citecolor=blue, linkcolor=blue}
\usepackage{color}
\usepackage{lipsum}

\newcommand{\ha}{H$\alpha$}
\newcommand{\hb}{H$\beta$}
\newcommand{\hg}{H$\gamma$}

\newcommand{\pab}{Pa$\beta$}
\newcommand{\brg}{Br$\gamma$}
\newcommand{\mbh}{\mbox{$M_\mathrm{BH}$}}
\newcommand{\rmin}{\mbox{$R_\mathrm{min}$}}
\newcommand{\rmean}{\mbox{$R_\mathrm{mean}$}}

\newcommand{\fvir}{\mbox{$f_\mathrm{virial}$}}
\newcommand{\sigl}{\mbox{$\sigma_\mathrm{line}$}}

\newcommand{\heii}{He{\sevenrm\,II}}
\newcommand{\hei}{He{\sevenrm\,I}}
\newcommand{\feii}{Fe{\sevenrm\,II}}

\newcommand{\OIIIb}{[O{\sevenrm\,III}]\,$\lambda$5007}
\newcommand{\OIIIc}{[O{\sevenrm\,III}]\,$\lambda\lambda$4959,5007}
\newcommand{\OIIId}{[O{\sevenrm\,III}]\,$\lambda$4363}
\newcommand{\OIII}{[O{\sevenrm\,III}]}

\newcommand{\NII}{[N{\sevenrm\,II}]}
\newcommand{\SII}{[S{\sevenrm\,II}]}
\newcommand{\HI}{H{\sevenrm\,I}}
\newcommand{\degree}{{\mbox{$^\circ$}}}
\newcommand{\kms}{{\mbox{$\mathrm{km\,s^{-1}}$}}}
\newcommand{\Msun}{\mbox{$M_\odot$}}
\newcommand{\vpeak}{\mbox{$v_\mathrm{peak}$}}

\font\sevenrm=cmr7 scaled 1000

\newcommand{\dybel}{\texttt{DyBEL}}

\newcommand{\mcc}[1]{\multicolumn{1}{c}{#1}}

\begin{document}

\title{
   Broad-line region geometry from multiple emission lines in a single-epoch spectrum
}

\author{
L.~Kuhn\inst{1,2} 
\and J.~Shangguan\inst{1,3} 
\and R.~Davies\inst{1} 
\and A.~W.~S.~Man\inst{2}
\and Y.~Cao\inst{1}
\and J.~Dexter\inst{1,4}
\and F.~Eisenhauer\inst{1,5}
\and N.~M.~F\"orster~Schreiber\inst{1} 
\and H.~Feuchtgruber\inst{1}
\and R.~Genzel\inst{1,6}
\and S.~Gillessen\inst{1}
\and S.~H\"onig\inst{7}
\and D.~Lutz\inst{1}
\and H.~Netzer\inst{8}
\and T.~Ott\inst{1}
\and S.~Rabien\inst{1}
\and D.~J.~D.~Santos\inst{1}
\and T.~Shimizu\inst{1}
\and E.~Sturm\inst{1}
\and L.~J.~Tacconi\inst{1}
}

\institute{
Max Planck Institute for Extraterrestrial Physics (MPE), Giessenbachstr.1, 85748 Garching, Germany\\
\email{shangguan@mpe.mpg.de}
\and Department of Physics \& Astronomy, The University of British Columbia, Vancouver, Canada BC V6T 1Z1
\and Kavli Institute for Astronomy and Astrophysics, Peking University, Beijing 100871, People’s Republic of China
\and Department of Astrophysical \& Planetary Sciences, JILA, University of Colorado, Duane Physics Bldg., 2000 Colorado Ave, Boulder, CO 80309, USA
\and Department of Physics, Technical University of Munich, 85748 Garching, Germany
\and Departments of Physics \& Astronomy, Le Conte Hall, University of California, Berkeley, CA 94720, USA 
\and Department of Physics and Astronomy, University of Southampton, Southampton, UK
\and School of Physics and Astronomy, Tel Aviv University, Tel Aviv 69978, Israel
}
             
\date{Received ...; accepted ...}

 
\abstract
{The broad-line region (BLR) of active galactic nuclei (AGNs) traces gas 
close to the central supermassive black hole (BH).  Recent reverberation mapping 
(RM) and interferometric spectro-astrometry data have enabled detailed 
investigations of the BLR structure and dynamics, as well as estimates of 
the BH mass.  These exciting developments motivate comparative investigations 
of BLR structures using different broad emission lines.  In this work, we have 
developed a method to simultaneously model multiple broad lines of the BLR from 
a single-epoch spectrum.  We apply this method to the five strongest broad 
emission lines (\ha, \hb, \hg, \pab, and \hei~$\lambda$5876) in the UV-to-NIR 
spectrum of NGC~3783, a nearby Type~I AGN which has been well studied by RM and 
interferometric observations.  Fixing the BH mass to the published value, we fit 
these line profiles simultaneously to constrain the BLR structure.  We find that 
the differences between line profiles can be explained almost entirely as being 
due to different radial distributions of the line emission.  We find that using 
multiple lines in this way also enables one to measure some important physical 
parameters, such as the inclination angle and virial factor of the BLR.  
The ratios of the derived BLR time lags are consistent with the expectation of 
theoretical model calculations and RM measurements.}


   \maketitle
%

\section{Introduction}

It is challenging to measure the mass of a supermassive black hole (BH) at 
the center of a galaxy, because ideally, one needs to resolve its sphere of 
influence \citep{Kormendy2013}.  This is true in the local Universe where 
dynamical methods are preferred.  At cosmic distances, one has to rely on 
scaling relations except when the broad-line region (BLR) of an active 
galactic nucleus (AGN) is observable.  The broad recombination lines with 
typical full width at half maximum (FWHM) $\gtrsim1000\,\kms$ 
\citep{Khachikian74} are emitted by the ionized gas surrounding the accreting BH 
\citep{Peterson1997}.  The BH mass can be derived with the virial method once 
the size of the BLR has been measured, 
$M_\mathrm{BH} = f_\mathrm{virial}R v^2/G$, where $R$ is the BLR radius; $v$ is 
a characteristic velocity of the BLR rotation; and \fvir\ is the virial factor 
which takes account of the geometry of the BLR.  The structure and dynamics of 
the BLR strongly affect the virial factor and are critical to the BH mass 
measurement \citep{Collin2006,MejiaRestrepo2018}.

The broad line profile suggests that the BLR has a disk-like geometry 
\citep[e.g.][]{Wills1986,Vestergaard2000,Kollatschny2011,Shen2014,StorchiBergmann2017}.
Its size is most often measured from the time lag between the AGN continuum and 
the broad emission line light curves, the reverberation mapping (RM) technique 
\citep{Blandford1982,Peterson2014}.  The characteristics of these time lags 
across different velocity channels provide evidence of inflow and outflow 
motions in the BLR \citep[e.g.][]{Bentz2010a,Grier2013,Du2016}.  This has led to 
the development of comprehensive models that can constrain the BLR structure 
using high-quality RM data \citep{Brewer2011,Pancoast2011,Li2013,Pancoast2014a,Pancoast2014b}.  
More recently, the BLR has been spatially resolved with spectro-astrometry (SA), 
which is a powerful technique for measuring the BLR structure and BH mass 
\citep{GC2018,GC2020,GC2021b,GC2023}.  Attempts have been made to analyze SA and 
RM data jointly (hereafter, the SARM method), to measure the geometric distance 
of the BLR and better constrain the BLR structure and BH mass \citep{Wang2020,GC2021a,Li2022}.

The high-quality data needed for the detailed analyses described above are not 
widely available for large AGN samples, even with the ongoing large RM 
projects, such as SDSS-RM \citep{Shen2015} and OzDES-RM \citep{Malik2023}.  
But there is a wealth of AGN samples with good quality single-epoch spectra.
To exploit these, \cite{Raimundo2019,Raimundo2020} modified the widely used 
BLR dynamical modeling code \texttt{CARAMEL}, and used it to fit single-epoch 
line profiles.  They were able to constrain some BLR model parameters, such as 
the inclination angle and disk thickness, and estimate a BH mass by setting a 
prior on the BLR size based on the empirical size--luminosity relation 
\citep[e.g.][]{Bentz2013}.  We turn this idea around and focus on investigating 
the BLR structure and the virial factors derived from multiple broad lines, 
which are covered simultaneously in one UV/optical/NIR spectrum.  By doing so, 
we can understand how the structure of the BLR changes between different lines 
within the same AGN.
Previous RM observations found that higher ionization lines respond more 
promptly to continuum variations than lower ionization lines
\citep[e.g.,][]{Clavel1991,Gaskell2009}.  The photoionization model, such as 
the ``locally optimally emitting cloud'' (LOC) model \citep{Baldwin1995}, can 
naturally produce such ``radial ionization stratification.''  \cite{Korista2004} 
predicted decreasing time lags of \ha, \hb, \hg, \hei, and \heii\ using the LOC 
model, which is confirmed by the RM observation of nearby AGNs \citep{Bentz2010b}.

In this paper, we analyze the VLT/X-Shooter \citep{Vernet2011} spectrum of 
NGC~3783, which covers several strong prominent broad emission lines of hydrogen 
and helium, and for which the high spectral resolution enables a robust 
decomposition of the broad and narrow lines.  These data were previously used in 
a study of BLR excitation and extinction in several AGN \citep{SchnorrMuller2016}.  
For NGC~3783, \cite{Bentz2021a} report the RM time lags of various lines 
including \hb\ and \hg.  \cite{Bentz2021b} performed dynamical 
modeling of \hb\ and \heii\ lines and derived a BLR size and BH mass 
consistent with the traditional method.  \cite{GC2021b} report the SA 
measurement of the broad \brg\ line.  This motivated a joint SARM 
analysis, which yielded consistent results \citep{GC2021a}.  Here, we introduce 
the data and our method to decompose the broad-line profiles in 
Section~\ref{sec:xsho}.  Then we make a nonparametric characterization of them 
in Section~\ref{sec:nonp}.  We discuss our modeling of the line profiles in 
Section~\ref{sec:model}, comparing our results with \cite{GC2021a} because 
the joint analysis provides the strongest constraint of the BLR model.  
We discuss the strengths and limitations of the current method in 
Section~\ref{sec:se} and conclude in Section~\ref{sec:con}.


\begin{figure*}
\begin{center}
\begin{tabular}{c c}
\multicolumn{2}{c}{\includegraphics[width=0.90\textwidth]{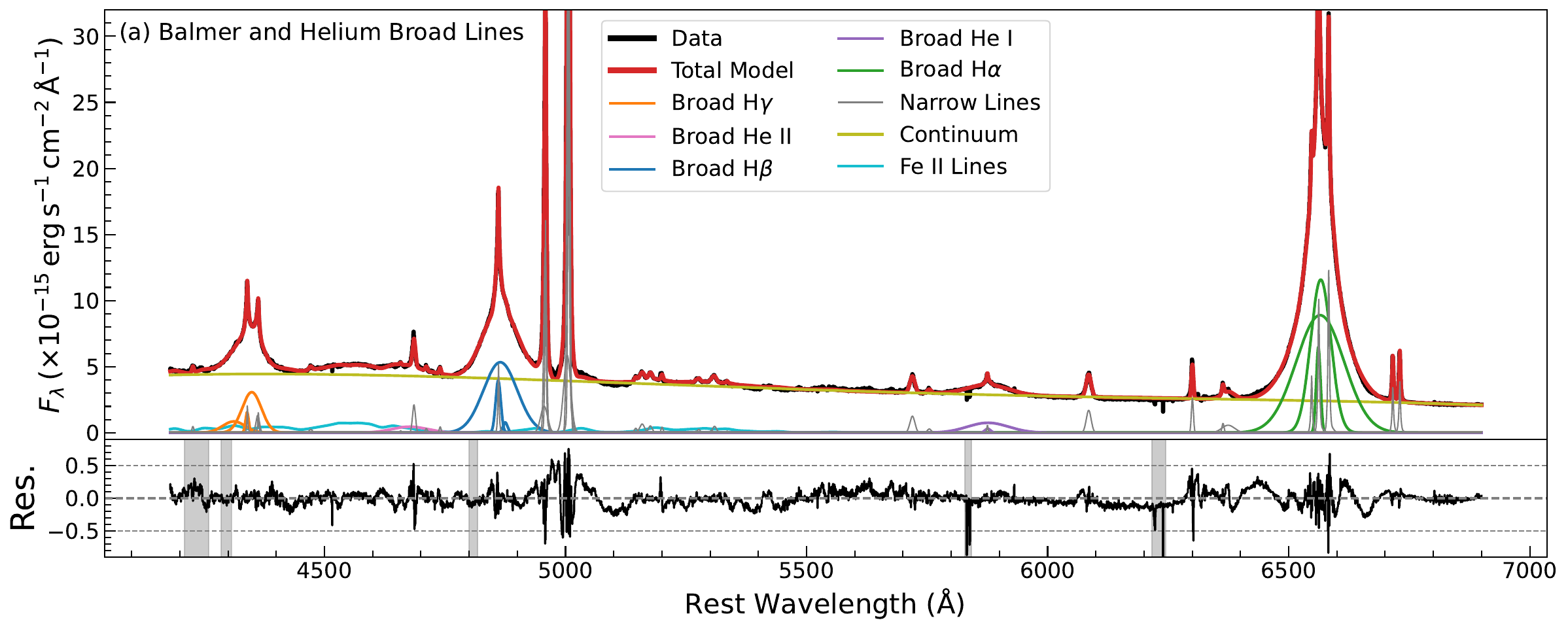}} \\
\includegraphics[width=0.53\textwidth]{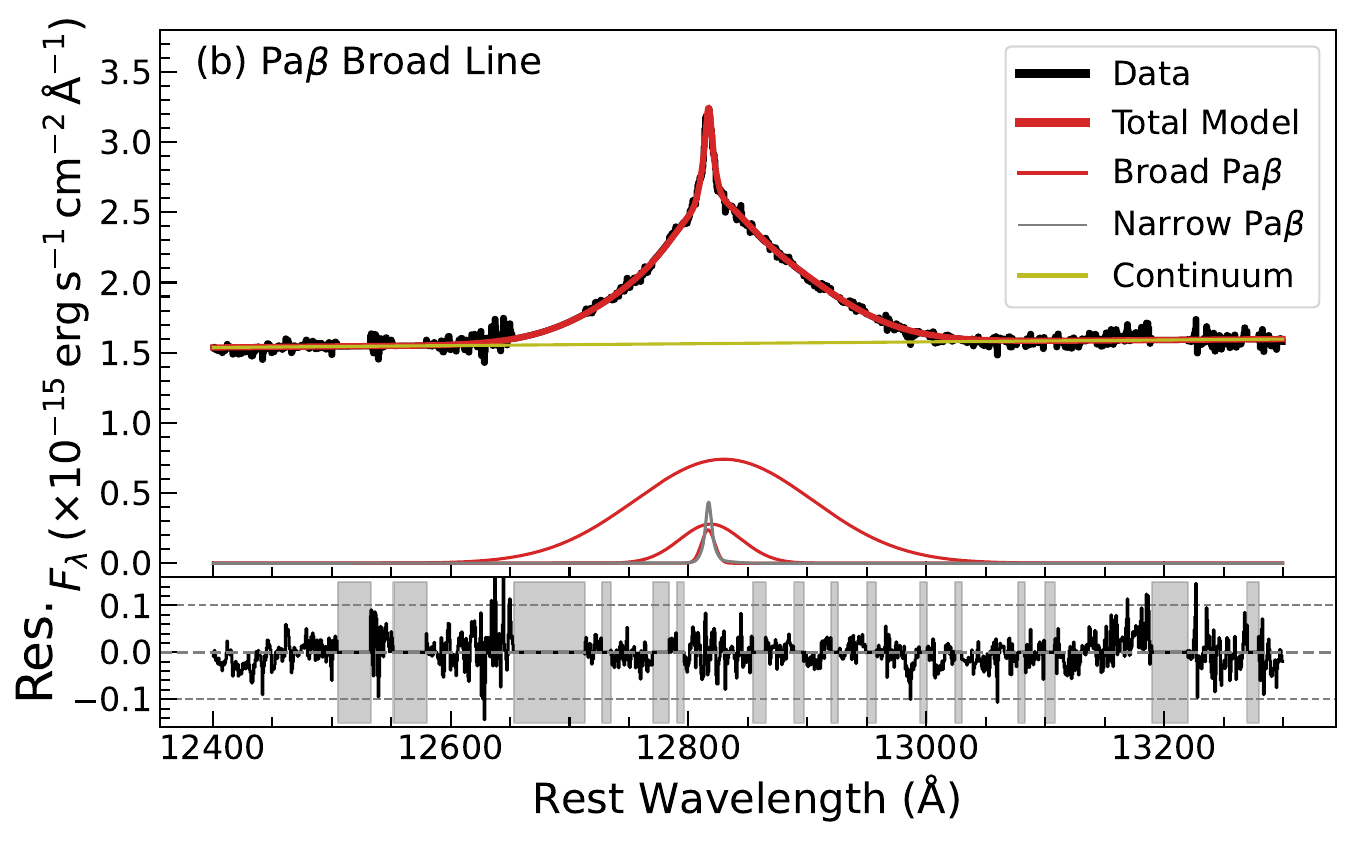} & \includegraphics[width=0.35\textwidth]{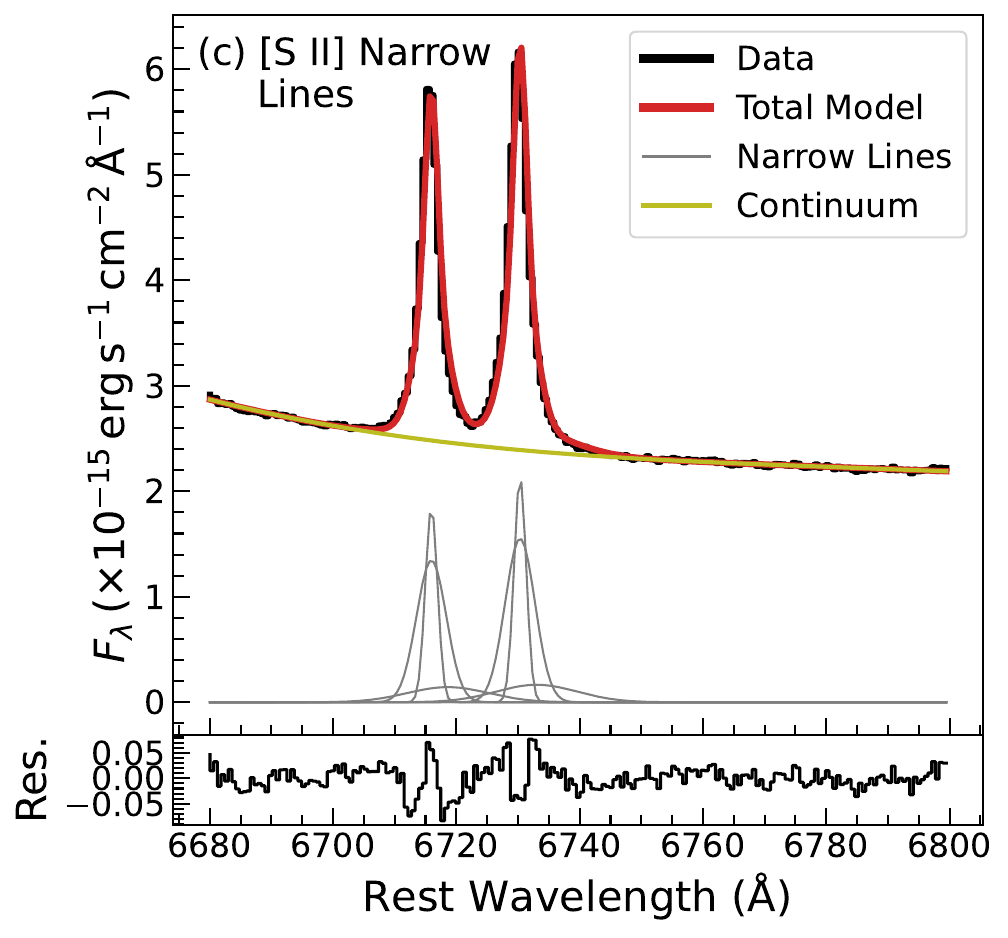} \\
\end{tabular}
\end{center}
\caption{X-shooter spectrum of NGC~3783 simultaneously covering UV, optical, and 
NIR ranges. In each panel, the spectrum is shown with the full model that enables 
decomposition of the broad line profiles overplotted, with the residual 
underneath.  The emission line components are also plotted separately. 
The shaded grey regions in the residuals represent wavelength ranges of bad 
channels and features due to the poor telluric correction, which are masked in 
the fitting.
(a) the UV/optical spectrum with the \ha, \hb, and \hg\ lines.  
(b) the part of the NIR spectrum used in this study, with the \pab\ line.  
(c) the \SII\ doublet used as the narrow line template.}
\label{fig:spec}
\end{figure*}

\section{Data reprocessing of X-shooter spectra}
\label{sec:xsho}

\subsection{Data reduction}
\label{ssec:data}

The X-shooter data were acquired as part of the LLAMA project \citep{Davies2015}. 
A description of the observations and data reduction can be found in 
\cite{SchnorrMuller2016} and \cite{Burtscher2021}.  We briefly summarize the key 
points as follows.  NGC~3783 (11:39:01.7, $-$37:44:19.0) was observed with 
X-shooter at the Very Large Telescope in early 2014, using the IFU mode 
(program ID 092.B-0083).  The spectral resolving power, 
$R = \lambda/\Delta \lambda$, is about 8400 (UVB), 13200 (VIS), and 8300 (NIR; 
\citealt{SchnorrMuller2016}).  The data were reduced with ESO reflex pipeline 
(version 2.6.8) with the Kepler GUI interface \citep{Modigliani2010} and mostly 
the default configuration.  The pipeline provides the data cubes of UVB, VIS, 
and NIR arms separately for each observation.  Telluric and flux calibrator stars 
were also observed.  Flux calibration was performed with a spectro-photometric 
standard from \cite{Moehler2014}.  From a comparison of the stars observed 
throughout the program, the spectrum is calibrated to an accuracy of about 2\%.  
We extracted 1D spectra from each of the NGC~3783 datacubes using a rectangular 
slit with a width of 1.8\arcsec, and applied minor scaling corrections to match 
the different spectral ranges.

We corrected Galactic extinction based on $A_V=0.332$ 
\citep{Schlafly2011} and the \cite{Cardelli1989} extinction model where we 
specify $R_V = 3.1$ as the ratio of total to selective extinction, and convert 
the spectrum to the rest frame adopting a redshift of 0.00973 measured from \HI\ 
21~cm line observations \citep{Theureau1998}.  A final correction was made to 
the blue wing of the broad hydrogen lines in the UV arm due to absorption in 
the telluric star.  We masked these narrow wavelength ranges (4286-4307$\AA$ and 
4800-4818$\AA$) when modelling the BLR profiles (Section~\ref{sec:model}).  
We also identify a few bad channels in the \hei~$\lambda$5876 profile and some 
channels contaminated by absorption and emission lines of the sky in the \pab\ 
profile.  We masked these channels when we modeled the line profiles, noting 
that they are always much narrower than the broad line profiles, so they will 
not affect the modeling.  We present the optical and NIR parts of the spectrum 
that are relevant to this work in Figure~\ref{fig:spec}, together with 
the spectral decomposition that will be discussed in Section~\ref{ssec:decmp}.

\begin{figure*}
\centering
\includegraphics[width=0.7\textwidth]{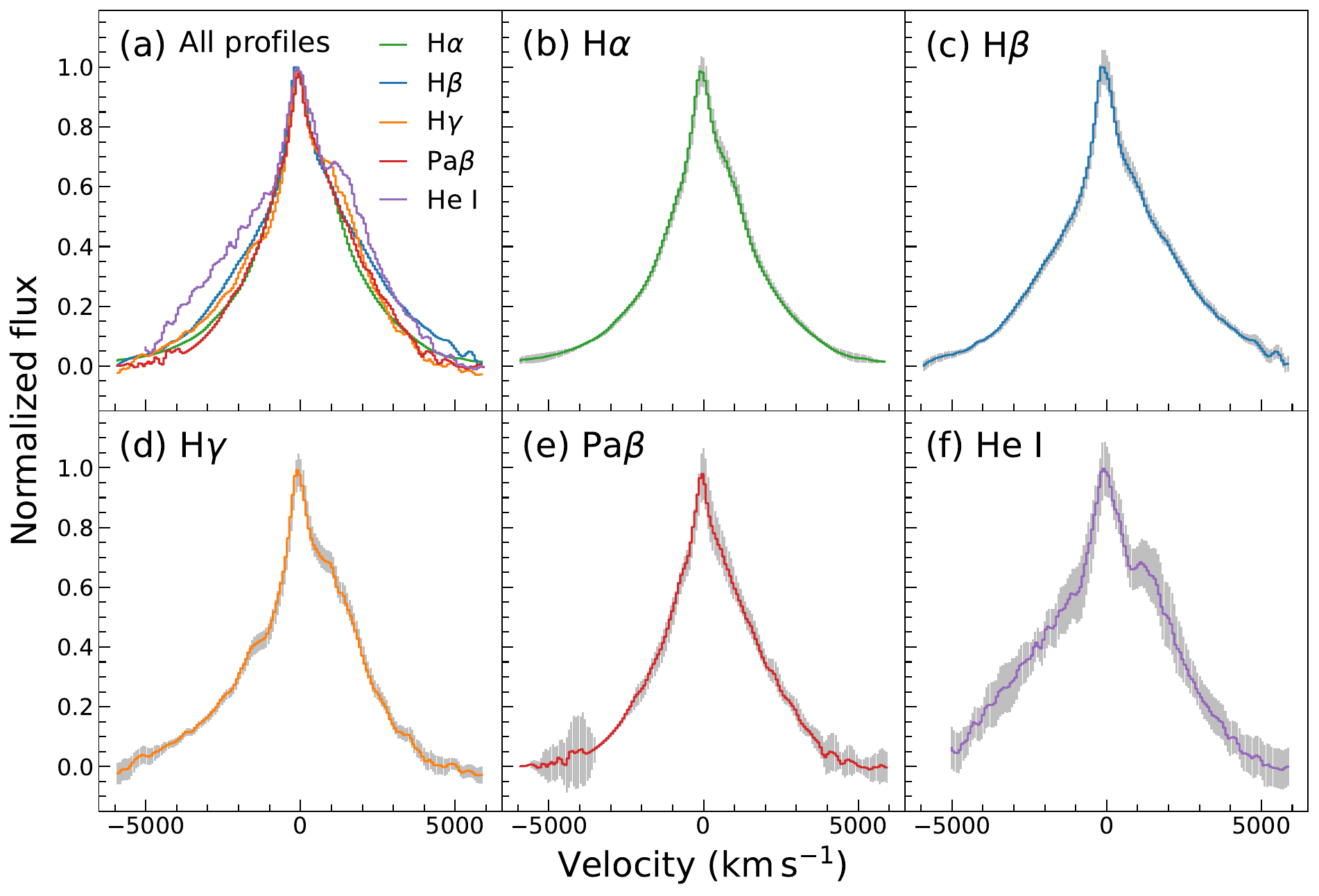}
\caption{Normalized broad emission line profiles of \ha, \hb, \hg, \pab, and \hei. 
These were extracted as described in Sec.~\ref{ssec:decmp}.  Panel (a) shows 
the profiles superimposed for an easier comparison, while panels (b)--(f) 
display the line profiles individually.  The uncertainties are shown in gray.  
The masked data are replaced by the multi-Gaussian model as shown in 
Figure~\ref{fig:spec} for clarity.}
\label{fig:bel}
\end{figure*}

\subsection{Spectral decomposition}
\label{ssec:decmp}

We decomposed the broad-line components of \ha, \hb, \hg, and \hei~$\lambda$5876 
from the combined UVB and VIS spectra, and the \pab\ profile from the NIR 
spectrum.  The final overall fit can be found in Figure~\ref{fig:spec}.

\subsubsection{Narrow line template}
\label{sssec:nlt}

The narrow line profiles of AGNs are usually more complex than a simple Gaussian 
profile, which argues for the use of a template based on isolated lines.  
The \SII~$\lambda\lambda$6716,~6731 and \OIII~$\lambda\lambda$4959,~5007 lines 
are usually used for this purpose, and one can use multiple Gaussian components 
to generate a noise less narrow line template \citep{Ho1997}.  To generate 
the template, we fit each of the \SII\ lines with 3 Gaussian profiles, tying 
their width and velocity shift from the laboratory wavelength between the pair, 
but allowing the total scaling to vary.  At the same time, we fit the local 
continuum with a 3rd-order polynomial function.  We found more Gaussian 
components or a higher-order polynomial function cannot improve the fitting.  
We have adopted the \SII\ doublet because the \OIII\ lines show stronger 
blue-shifted wind components: when trying a template based on \OIII, we found it 
did not match the narrow line components of the \HI\ lines well.

\subsubsection{Decomposing the broad-line profiles}
\label{sssec:bel}

The continuum of an AGN optical spectrum consists of emission from the accretion 
disk, the host galaxy, as well as the pseudo-continuum of the \feii\ lines 
\citep[e.g.][]{Barth2015}.  We adopted a power-law model for the AGN featureless 
continuum and found a host galaxy component based on \cite{Bruzual2003} stellar 
populations amounted to $\lesssim 10\%$ of the continuum and was thus 
not needed. To fit the \feii\ features, we incorporated the newly published 
high-quality template covering 4000--5600~\AA\ \citep{Park2022}. 
We found the final normalized line profiles only change $\lesssim 0.05$ 
if we adopt the \feii\ template from \cite{Boroson1992}, which has little effect on  
the BLR modeling results.  We broadened and shifted the \feii\ template as 
part of the fitting process.  A wide wavelength range is useful for decomposing 
these components, so we opted to fit the entire optical spectrum over 
4200--6800~\AA\ and decompose the broad \ha, \hb, \hg, and \hei\ simultaneously 
(Figure~\ref{fig:spec}a).  Fitting the more isolated the \pab\ line is described 
later.

In the optical spectrum, the majority of narrow lines are fitted by the \SII\ 
template with two free parameters, the amplitude and the velocity shift from 
the laboratory wavelength.  We tie the velocity shifts of all the templates, 
so it reflects a small deviation ($\approx -2\,\kms$) from the redshift measured 
by the atomic \HI\ gas.  Because the UVB arm has a lower spectral resolution 
than the VIS arm, we broaden the narrow line template with a Gaussian kernel 
($\sigma \approx 35\,\mathrm{km\,s^{-1}}$) for all lines at wavelengths shorter 
than 5600~\AA.  For the \OIII\ lines, which have a higher critical density and 
more contribution from blue-shifted components than \SII, we add 
additional Gaussian components: one for \OIIId, and two (tied together, and with 
a ratio of 2.98) 
for each line in the \OIIIc\ doublet.  The \NII~$\lambda$6550,~6585 doublets are 
fitted with the \SII\ template with the amplitude ratio fixed to the theoretical 
value of 2.96.  For completeness and to avoid influencing the continuum 
placement, we fitted several other narrow lines\footnote{We include 
[Fe~V]~$\lambda$4227, [O~III]~$\lambda$4363, He~I~$\lambda$4471, 
[Fe~III]~$\lambda$4658, [Ar~IV]~$\lambda$4711, [Ar~IV]~$\lambda$4740,  
[Fe~VI]~$\lambda$5146, [Fe~VII]~$\lambda$5159, [Fe~VI]~$\lambda$5176, 
[N~I]~$\lambda$5200, [Fe~III]~$\lambda$5270, [Fe~VII]~$\lambda$5276, 
[Ca~V]~$\lambda$5309, [Fe~XIV]~$\lambda$5303, [Fe~VI]~$\lambda$5335, 
[Fe~VII]~$\lambda$5721, [N~II]~$\lambda$5755, [Fe~VII]~$\lambda$6087, 
[O~I]~$\lambda$6300, [O~I]~$\lambda$6364, and [Fe~X]~$\lambda$6375.  While we 
fit the [O~III]~$\lambda$4363 with the narrow line template, we simply adopt 
a Gaussian function for the remaining narrow lines listed above.} in 
the spectrum, although they do not directly affect the broad line decomposition. 

For the broad lines, we found three Gaussian components are sufficient to fit 
the \ha, \hb, \hg, and \hei\ profiles.  We fit the broad \heii~$\lambda$4686 as 
well, although it is too faint to provide a robust line profile for our 
dynamical modeling.

Lastly, we multiplied the entire optical spectrum model by a 5th-order 
polynomial function to account for large-scale variations due to 
the instrumental and calibration effects \citep{Cappellari2017}.  This method 
can moderately improve the fitting of the continuum in the line wings at a level 
of $\sim 10^{-16}\,\mathrm{erg\,s^{-1}\,cm^{-2}\,\AA^{-1}}$.

In the NIR spectrum, we fitted the broad \pab\ profile and continuum over 
a wavelength range of 12400--13300~\AA.  We included a power-law for 
the continuum, the \SII\ template (broadened to match the resolution of the NIR 
arm) for the narrow \pab\ line, and three Gaussian components for the broad 
\pab\ line.  The velocity shift with respect to the theoretical wavelength of 
the narrow \pab\ line is not tied to the optical narrow lines.  But their 
velocity difference of $\lesssim 1.7~\kms$ is consistent with the systematic 
uncertainty of X-shooter wavelength calibration over different 
arms.\footnote{\url{https://www.eso.org/sci/facilities/paranal/instruments/xshooter/doc/XS_wlc_shift_150615.pdf}}  
This high accuracy enables us to tie the central wavelength offsets when we fit 
the broad line profiles simultaneously (Section~\ref{sec:model}).

\subsubsection{Resampling and uncertainties}
\label{sssec:res}

While the high resolution of the X-shooter spectrum helps to decompose 
the narrow and broad line profiles robustly, it is not necessary for 
the modeling which is only constrained by the broad 
($\gtrsim 1000\,\kms$) and smooth features.  Therefore, we resampled 
the decomposed line profiles to an effective resolving power of 2000, which is 
high enough to retain the characteristic features of the profiles while reducing 
the computational time.  We also verified that our conclusions do not change 
with slightly lower ($R=1000$) or higher ($R=4000$) resolution.

We first convolved the decomposed broad line profiles with Gaussian kernels to 
the required resolution.  Then, we used the Python tool \texttt{SpectRes} 
\citep{Carnall2017} to resample them, while preserving the integrated flux and 
propagating the uncertainties.  We chose the channel width of the re-sampled 
profiles to be $\sim 75\,\kms$ so that the spectra are still Nyquist sampled.

To calculate the uncertainties of the line profiles, we summed two components in 
quadrature: (1) the uncertainty of the observed spectrum and the line profile 
decomposition, and (2) 5\% of the line flux.  To estimate the first component, 
we used a running root mean square (RMS) of the fitting residual of the original 
spectra, which includes the imperfectness of the fitting and other potential 
artifacts.  The second component is important to avoid too much emphasis on 
the line center of the strongest lines (i.e. \ha\ and \hb).  Specifically, we 
adopt the following equation,
\begin{equation}
F_{\mathrm{err}} = \sqrt{\left(\mathrm{RMS}_{\mathrm{run}}(\mathrm{res})\times\sqrt{N_\mathrm{org}/N_\mathrm{dwn}}\right)^2+ (0.05F_\lambda)^2}
\end{equation}
where $\mathrm{RMS}_{\mathrm{run}}(\mathrm{res})$ is the running 
root-mean-squared (RMS) over the spectrum residuals with a window size of 30, 
$N_\mathrm{org}/N_\mathrm{dwn}$ is the ratio of the number of channels in 
the original spectrum over the downgraded spectrum,\footnote{we re-bin 
the spectrum profiles according to ratios of $\sim$2.26 and 3.32 for arms VIS 
and UVB/NIR, respectively} and $F_\lambda$ is the flux density.  

As a final step, we normalize the line profiles and their uncertainties 
according to the peak of our multi-Gaussian model of the broad lines in 
Section~\ref{sssec:bel}.  The resulting profiles are shown in 
Figure~\ref{fig:bel}.  The centroid and width of the line peaks are well 
consistent, while the width of the line wings varies for different lines. 
We note that the red wing of the broad \hg\ overlaps with the \OIIId\ line, 
which is fitted with the narrow line template plus a Gaussian component 
(Section~\ref{sssec:bel}).  We opted to keep the \OIIId\ model simple to avoid 
biasing the \hg\ profile.  However, this results in a relatively large residual 
(0--2000 \kms), and therefore uncertainty, of the line profile as shown in 
Figure~\ref{fig:bel}(d). The \hg\ line shows the strongest asymmetry due to 
the ``shoulder'' on the red wing, which is very difficult to model.  We believe 
it can be at least partly explained in terms of the decomposition.  In addition, 
the entire broad \hei\ profile shows relatively large uncertainties because this 
line is very weak compared to the \HI\ lines, and its uncertainties are dominated 
by the RMS term.  \hei\ is the most susceptible to any artifact of the spectrum 
among the five broad lines studied in this work.

\begin{table*}[]
\begin{center}
\caption{Nonparametric properties of the broad line profiles}
\label{tab:line}
\renewcommand{\arraystretch}{1.2}
\begin{tabular}{crrrrrrrr}
\hline\hline
Line  & \mcc{$\lambda_\mathrm{air}$} & \mcc{$v_\mathrm{peak}$} & \mcc{$W_{25}$} & \mcc{$W_{50}$} & \mcc{$W_{75}$} & \mcc{$\sigma_\mathrm{line}$} & \mcc{A.I.} & \mcc{K.I.} \\
      & \mcc{($\AA$)} & \mcc{($\mathrm{km\,s^{-1}}$)} & \mcc{($\mathrm{km\,s^{-1}}$)} & \mcc{($\mathrm{km\,s^{-1}}$)} & \mcc{($\mathrm{km\,s^{-1}}$)} & \mcc{($\mathrm{km\,s^{-1}}$)} & & \\
(1)   & \mcc{(2)} & \mcc{(3)}   & \mcc{(4)}    & \mcc{(5)}    & \mcc{(6)}    & \mcc{(7)}       & \mcc{(8)}     & \mcc{(9)}       \\\hline
 \ha  & $6562.8$  & $-70\pm81$  & $4360\pm113$ & $2371\pm112$ & $930\pm124$  & $1815\pm14$ & $0.18\pm0.08$ & $0.21 \pm 0.03$ \\ 
 \hb  & $4861.3$  & $-139\pm91$ & $5475\pm150$ & $2534\pm183$ & $875\pm116$  & $2002\pm15$ & $0.25\pm0.09$ & $0.16 \pm 0.02$ \\ 
 \hg  & $4340.5$  & $-62\pm79$  & $4784\pm137$ & $2549\pm138$ & $828\pm145$  & $1682\pm33$ & $0.37\pm0.07$ & $0.17\pm0.03$   \\ 
 \pab & $12818.1$ & $-2\pm94$   & $4706\pm207$ & $2516\pm215$ & $955\pm231$  & $1619\pm71$ & $0.17\pm0.09$ & $0.20\pm0.04$   \\ 
 \hei & $5875.6$  & $-60\pm118$ & $6374\pm316$ & $3542\pm388$ & $1117\pm225$ & $1954\pm42$ & $0.16\pm0.13$ & $0.18\pm0.03$   \\
\hline
\end{tabular}
\end{center}
{\small \textbf{Note.} 
Column (1): Line name; 
Column (2): Laboratory wavelength of the line in the air;
Column (3): Peak velocity of the broad line in the rest frame w.r.t. the $\lambda_\mathrm{air}$;
Column (4)--(6): The line widths at 25, 50, and 75 percentiles of the line peak;
Column (7): The second moment of the line profile, following the definition of 
the Equation (3) of \cite{DallaBonta2020};
Column (8): The asymmetry index (Equation~\ref{eq:ai});
Column (9): The kurtosis index (Equation~\ref{eq:ki}).
}
\end{table*}

\begin{figure*}
\centering
\includegraphics[width=0.9\textwidth]{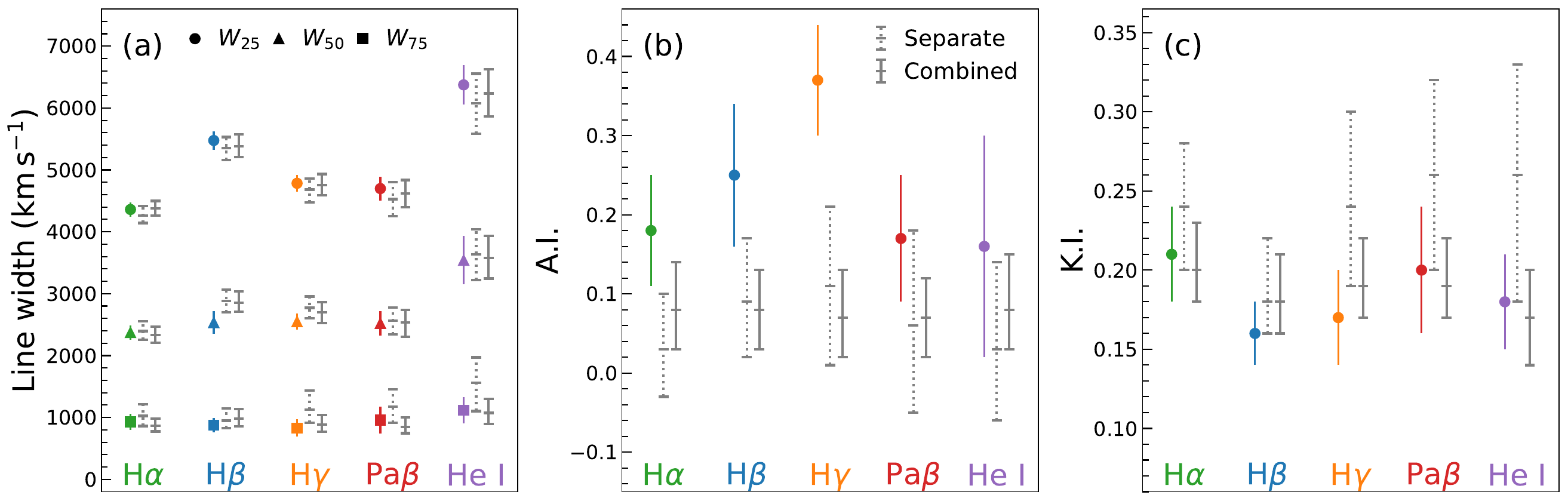}
\caption{Comparison of the nonparametric properties of the line profiles 
(denoted by different colors, and with 1$\sigma$ uncertainties). (a) line widths 
(circles, triangles, and squares represent $W_{25}$, $W_{50}$, and $W_{75}$, 
respectively); (b) asymmetry index (A.I.); (c) kurtosis index (K.I.).  The gray 
lines correspond to equivalent measurements on the model line profiles for 
the separate (dotted) and combined (solid) fitting results.}
\label{fig:npar}
\end{figure*}

\section{Nonparametric properties of the broad line profiles}
\label{sec:nonp}

Because broad-line profiles cannot be described by a simple analytical function, 
we characterize them nonparametrically and later use these quantities 
(Table~\ref{tab:line}) to assess the validity of our model.  The peak velocity, 
\vpeak, is the deviation of the peak wavelength of the line from its expected 
wavelength, after shifting to the rest frame as described in 
Section~\ref{ssec:data}.  All of the broad lines peak near the systemic velocity.  
We calculate the Full Width at 25\% ($W_{25}$), 50\% (FWHM, $W_{50}$), and 75\% 
($W_{75}$) Maximum.  We also calculate the Asymmetry Index (A.I.) and Kurtosis 
Index (K.I.) of each line profile as defined in \cite{Marziani1996},
\begin{align}
\mathrm{A.I.} &= \frac{v_R(1/2) + v_B(1/2) - 2\vpeak}{v_R(1/2) - v_B(1/2)}, \label{eq:ai}\\
\mathrm{K.I.} &= \frac{v_R(3/4) - v_B(3/4)}{v_R(1/4) - v_B(1/4)}, \label{eq:ki}
\end{align}
where $v_R(x) > 0$ and $v_B(x) < 0$ ($x=1/4, 1/2,\, \mathrm{or}\, 3/4$) are 
the velocity of line profiles at the corresponding fraction of the line peak 
on the red and blue wings respectively.  The values of A.I. indicate 
the direction and degree of asymmetry in the line profile shape.  
Differing slightly from the definition of \cite{Marziani1996}, we calculate 
the A.I. at 50\% (instead of 25\%) of the peak flux because our line profiles 
become more symmetric towards the wings.  A positive A.I. indicates that 
the line profiles are skewed towards the red side relative to the profile 
center.  This suggests an excess of line emission or broader velocity 
distribution on the red side compared to the blue side.   K.I., on the other 
hand, is essentially $W_{75}/W_{25}$.  A Gaussian profile has 
K.I.~$\approx 0.46$.  A smaller K.I. indicates that the line profile has 
a broader wing than a Gaussian profile, and vice versa.  We also 
calculate the second moment of the line profiles, $\sigma_\mathrm{line}$, 
following the definition of the Equation (3) of \cite{DallaBonta2020} for 
completeness because $\sigma_\mathrm{line}$ is widely used in RM works to derive 
the BH mass \citep{Peterson2004,Wang2019}.  To determine the uncertainty for 
each reported value, we randomly perturb each profile using the uncertainties 
from Section~\ref{sssec:res}, re-measure these quantities 500 times, and 
calculate the standard deviation of the results.

We compare these parameters in Figure~\ref{fig:npar}.  The \HI\ lines show 
almost the same widths at 75\%  and 50\% of the line peaks.  The \ha\ 
line shows the lowest $W_{25}$, while \hb\ shows the highest $W_{25}$ among 
the \HI\ lines.  The line wing of \hb\ is significantly broader than the other 
\HI\ lines (see also Figure~\ref{fig:bel}).  Although the \hb\ region is 
complicated, the line wing cannot be biased by the spectral decomposition.  
As shown in Figure~\ref{fig:spec}, the broad \hb\ line is much stronger than 
the (pseudo-)continuum and far enough from the \OIII\ lines.  The \hei\ line 
shows a comparable line width to the \HI\ lines at its peak ($W_{75}$) but 
becomes much broader towards the wing ($W_{50}$ and $W_{25}$).  
The $\sigma_\mathrm{line}$ of all lines are similar, and in between 
their $W_{50}$ and $W_{75}$, respectively.  We do not plot them in 
Figure~\ref{fig:npar} for clarity.  All the lines show measurable and positive 
A.I.; and although the Balmer lines show differing values, there are large 
uncertainties.  The K.I. values are similar among the lines, and are smaller 
than that expected for a single Gaussian profile, indicating that the profiles 
are peaky with broad wings.

\section{Modeling the broad-line profiles}
\label{sec:model}

In this section, we first introduce our BLR dynamical model, 
its limitations, and our inference 
strategy (Section~\ref{ssec:dybel}).  We then model the broad-line profiles in 
two steps: 
(1) We fit the line profiles separately and study the consistency of the model 
parameters (Section~\ref{ssec:sep}); 
(2) We fit the line profiles with almost all the BLR parameters tied and only 
allow the radial distribution of different line emissions to vary freely 
(Section~\ref{ssec:sim}).

\begin{table*}[]
\begin{center}
\caption{Summary of the BLR model parameters, including a short explanation and 
prior range}
\label{tab:panc}
\renewcommand{\arraystretch}{1.2}
\begin{tabular}{ccc}
\hline\hline
Parameter & Explanation & Priors \\
\hline
$i$                 & Inclination angle                                                          & Uniform($\cos 60^\circ$, $\cos 0^\circ$) \\ 
$\theta_o$          & Angular thickness measured from the mid-plane                              & Uniform($0^\circ$, $90^\circ$) \\ 
$f_\mathrm{ellip}$  & Fraction of clouds in bound elliptical orbits                              & Uniform(0, 1) \\
$\kappa$            & Anisotropy of the cloud emission                                           & Uniform($-0.5, 0.5$) \\
$\gamma$            & Clustering of the clouds at the edge of the disk                           & Uniform(1, 5)\\
$\xi$               & Mid-plane transparency                                                     & Uniform(0, 1)\\
$\theta_\mathrm{e}$ & Angular location of the radially moving clouds in ($v_r$, $v_\phi$) space  & Uniform($0^\circ$, $90^\circ$) \\
$\mu$               & Mean radius of cloud distribution                                          & LogUniform($10^{-4}$~pc, 10~pc) \\
$\beta$             & Unit standard deviation of BLR radial profile                              & Uniform(0, 2) \\
$F$                 & Ratio of the minimum cloud radius and the mean radius                      & Uniform(0, 1) \\
$\epsilon$          & Central wavelength offset, $\epsilon = \lambda_c/\lambda_\mathrm{air} - 1$ & Normal($0, 10^{-4}$) \\
$f_\mathrm{peak}$   & Peak flux of the normalized line profile                                   & Normal($1, 10^{-2}$) \\
$f_\mathrm{flow}$   & Flag for specifying inflowing or outflowing orbits                         & Fixed($<0.5$) \\
$M_\mathrm{BH}$     & Black hole mass                                                            & Fixed($10^{7.4}\,M_\odot$) \\
\hline
\end{tabular}
\end{center}
\end{table*}

\subsection{BLR dynamical model and the inference}
\label{ssec:dybel}

The nature of the BLR is still an open question, and many models have 
been proposed to explain its various aspects 
\citep[][and references therein]{Peterson2006,Czerny2019}. One major class of 
models assumes that the BLR consists of many discrete clouds 
\citep[e.g.][]{Rees1989,Baldwin1995,Czerny2011,Baskin2018,Rosborough2023}. 
The cloud model offers advantages in parameterizing the geometry, kinematics, 
and photoionization physics flexibly, enabling interpretation of observations, 
in particular recent high-quality RM and interferometric data 
\citep{Korista2004,Pancoast2011,Pancoast2014a,Li2013,Li2018,Williams2022,GC2018,GC2020,GC2021b}. 
However, the physics of the cloud model may be oversimplified. For example, 
there is ongoing debate regarding the confinement of gas within high-density 
($\gtrsim 10^9\,\mathrm{cm}^{-3}$) clouds 
\citep{Mathews1986,Rees1987,Krolik1988,Baskin2014,Proga2014,Proga2015}. 
Moreover, as discussed in detail by \cite{Netzer2020}, state-of-the-art cloud 
photoionization models tend to underproduce the luminosity of the Balmer lines 
(and other non-resonant \HI\ lines) by a factor of 2--5, likely due to 
the failure of escape probability formalism in photoionization codes like CLOUDY 
\citep{Ferland1998} for high densities and optical depths in the BLR. Radiation 
hydrodynamic simulations of the disk wind, coupled with radiative transfer 
calculations, have been deemed crucial to understand the photoionization physics 
of the BLR \citep{Waters2016,Matthews2016,Matthews2020,Mangham2017}. However, 
the high computational expense impedes the development of more comprehensive 
models for detailed data interpretation.

In this work, we employ our self-implemented BLR dynamical model to 
characterize the distribution and kinematics of the BLR line emission (or 
`emissivity'). The model parameterization was initially developed in 
\cite{Pancoast2014a}. Our model has been utilized to fit the normalized line 
profile and differential phase signal, tracking the spatially resolved 
kinematics of recent interferometric observations of BLRs 
\citep{GC2020,GC2021a,GC2021b}. Where our implementation differs from \cite{Pancoast2014a}, is in using 
a Monte Carlo cloud model to depict the line emission at the moment of the observation, excluding variable continuum light curves and reverberation mapping 
physics. Thus, our modeling approach circumvents the challenges associated with 
the aforementioned photoionization physics of the BLR by modeling 
the line emission distribution instead of the physical clouds and 
photoionization physics. In making this statement, we emphasize 
that our use of the term `cloud' should be taken to mean `line emitting 
entity'. It does not refer to physical `gas clouds,' nor 
does it indicate a preference for the cloud model over the so-called disk-wind 
model \citep[e.g.][]{Chiang1996,Matthews2016,Long2023}. Here, we adapted the BLR 
model into a Python package, \dybel, allowing the fitting of line profiles 
exclusively. \dybel\ can be used to fit either a single line profile or multiple 
lines of an AGN simultaneously. As the detailed BLR model is presented in 
\cite{GC2020}, we provide a brief introduction to the key parameters, summarized 
in Table~\ref{tab:panc}.

The model comprises a large number\footnote{We adopt $10^5$ clouds in 
the fitting which is large enough to produce smooth line profiles.} of 
non-interacting mass-less point particles orbiting the central BH with mass 
\mbh, forming a disk-like structure.  The radial distribution of the clouds 
follows a shifted Gamma function governed by three parameters: $\mu$ as the mean 
radius, $F=\rmin/\mu$ with \rmin\ the minimum cloud radius, and $\beta$ for 
the shape of the profile (Gaussian: $0<\beta<1$, exponential: $\beta=1$, and 
heavy-tailed: $1<\beta<2$).  The angular thickness of the disk is $\theta_o$, 
and the vertical distribution of the clouds is governed by $\gamma$, with a 
higher value ($\gamma>1$) corresponding to more clouds concentrating on the disk 
surface.  The structure is viewed at an inclination angle $i$ (with 0\degree\ 
corresponding to a face-on view).  Each cloud is randomly assigned to be on 
a quasi-circular orbit with radial and tangential velocities ($v_r$, $v_\phi$) 
around (0, $v_\mathrm{circ}=\sqrt{G\mbh/r}$), or a quasi-radial orbit.  
The fraction on quasi-radial orbits is controlled by $f_\mathrm{ellip}$ 
(with $f_\mathrm{ellip}=1$ meaning all clouds are on such orbits).  
A binary parameter, $f_\mathrm{flow}$, governs the direction of the cloud 
radial motion.  These clouds are inflowing if $f_\mathrm{flow}<0.5$, or 
outflowing if $f_\mathrm{flow}>0.5$.  Their radial and tangential velocities are 
controlled by an angular parameter $\theta_e$ such that when $\theta_e=0$ 
the velocity vector is 
($v_\mathrm{esc}=\sqrt{2}v_\mathrm{circ}$, 0), and when $\theta_e=\pi/2$ it is 
(0, $v_\mathrm{circ}$).  While \cite{Pancoast2014a} had additional parameters 
defining how the cloud velocities are dispersed around these points, we exclude 
them because they are generally unconstrained and do not influence our fitting 
\citep{GC2020,GC2021a,GC2021b}.  Finally, the weight of each cloud is 
controlled by $-0.5<\kappa<0.5$, reflecting the anisotropy of the cloud 
illumination.  Clouds closer to the observer have higher weights if $\kappa>0$ 
and vice versa.  The ratio of clouds below and above the midplane is controlled 
by $\xi$, reflecting the ``midplane obscuration'' of the BLR.  There are equal 
amounts of clouds between the midplane if $\xi=1$, while there is no cloud below 
the midplane if $\xi=0$.  To fit the line profiles, we need two nuisance 
parameters, the central wavelength ($\lambda_c$) and the peak flux 
($f_\mathrm{peak}$) of the line.

We use the above BLR model to describe the line emission distribution of 
the BLR without including any photoionization physics.  
We cannot predict the physical line luminosity with this model, so the line peak 
flux is a free parameter in the fitting and we only model the normalized line 
profiles.  This approach is adopted by the recent works of 
\cite{GC2018,GC2020,GC2021b,GC2023} when they model the line profile and 
differential phase data (equivalent to spatially resolved kinematic data).  
In contrast, the original application of this model to the RM data assumes 
the point particles as `mirrors' reflecting the continuum 
emission, the limitations of which are nicely summarized by 
\citet[][in their Section 2.2]{Raimundo2020}.  Their arguments make it clear 
that photoionization physics is needed in the application of RM modeling.  
Indeed, there is recent progress in addressing this problem \citep{Williams2022,Rosborough2023}.  
Nevertheless, the recent study of NGC~3783 shows that the modeling using GRAVITY 
and RM data is remarkably consistent \citep{Bentz2021b,GC2021a,GC2021b}, which 
supports the application of this simple model at least for the particular case of 
NGC~3783.  We discuss the caveats of applying this model to the single-epoch 
spectra, which is the main goal of this work, in Section~\ref{ssec:cav}.

We use the Python package \texttt{dynesty} \citep{Speagle2020} to fit the data 
with a nested sampling algorithm, which is more powerful than the typical Markov 
Chain Monte Carlo method for complex models with many (e.g. $>20$) parameters 
and a potentially multimodal posterior distribution.  By design, the nested 
sampling algorithm \citep{Skilling2004} can estimate the Bayes evidence, which 
enables us to compare different models.  We use the dynamic nested sampler 
(\texttt{DynamicNestedSampler}) which better estimates the likelihood function 
by re-sampling the posterior function a few more times after 
the ``baseline run.''  We use 1200 live points for the baseline run and add 500 
points for each of the 10 re-samplings.  We adopt the random walk algorithm 
(\texttt{rwalk}) to sample the prior space and use the multi-ellipse method 
(\texttt{multi}) to create the nest boundaries.  We adopt the default values for 
all the remaining options of \texttt{dynesty}.

The metric we use to define the goodness of fit is the likelihood function, 
\begin{align}
\ln\,\mathcal{L} = -\frac{1}{2} \sum^N_{l} \sum^n_{i} \left(\frac{(f_{l,i} - \tilde{f}_{l,i}(\vec{\Theta}_l))^2}{\sigma^2_{l,i} T} \right),
\end{align}
where the first summation over $l$ is for different lines and the second 
summation over $i$ is for different channels of a line profile; $f_{l,i}$ and 
$\sigma_{l,i}$ are the line profile flux and uncertainty and 
$\tilde{f}_{l,i}(\vec{\Theta}_l)$ is the corresponding line model with the set 
of parameters $\vec{\Theta}_l$; and a temperature parameter, $T>1$, is included 
to effectively scale up the uncertainties of the data.  The temperature makes 
the likelihood function less peaky, which facilitates proper estimation of 
the posterior distribution.  We found that $T = 16$ is a suitable setting, and 
our fitting results are not sensitive to the specific choice of temperature 
(e.g. $T=8$, 16, or 32).

It is important to bear in mind that when fitting only the line profiles, \mbh\ 
and $\mu$ are fully degenerate.  This is because the cloud velocities always 
scale with $v_\mathrm{circ}$ which depends on $\mbh/r$, and it means that one 
needs to fix the BH mass in order to investigate the BLR sizes.  Therefore, we 
include physical prior information by fixing $\mbh = 10^{7.4}\,\Msun$ 
\citep{GC2021a}.  The exact value of the \mbh\ does not affect 
the derived BLR model parameters except the BLR radius.  We will discuss this 
point in more detail in the following sections.  Next, we set 
$f_\mathrm{flow}$, a binary flag to decide the direction of the radial velocity 
of the clouds.  While the previous modeling efforts all indicate radial inflow 
\citep{Bentz2021b,GC2021a,GC2021b}, using line profiles alone cannot distinguish 
between inflow and outflow (either in the model or via the Bayes evidence) 
because there is no spatial information, and the specific choices of 
$f_\mathrm{flow}$ do not affect our results.  Therefore, we adopt an inflow 
model setting $f_\mathrm{flow}<0.5$.

When we fit the line profiles simultaneously with \dybel, we can choose to tie 
additional parameters besides fixing the same BH mass.  Our aim is to assess 
whether the difference in the line profiles can be attributed to radial 
differences in the line emission distribution for an otherwise fixed BLR 
geometry.  We therefore leave $\mu$, $F$, and $\beta$ free to vary for each 
line, while the remaining model parameters are tied.  We also tie the central 
wavelength offsets for each line, allowing them to shift together by a small 
amount $\epsilon = \lambda_c/\lambda_\mathrm{air} - 1$ where 
$\lambda_\mathrm{air}$ is the nominal wavelength in air.  We adopt a Gaussian 
prior centered at 0 with a small standard deviation of 0.01 for $\epsilon$ 
because $\lambda_c$ is expected to be close to $\lambda_\mathrm{air}$.  
Similarly, the normalized line peaks are expected to be close to 1, so we only 
adopt a single nuisance parameter, $f_\mathrm{peak}$, in the fitting with 
a Gaussian prior centered at 1 with a standard deviation of 0.1.  The priors of 
the remaining parameters are adopted from \cite{GC2020}.  

For comparison, we first fit each profile separately, then mainly discuss 
the results fitting all of the five lines simultaneously.  In each case, we 
calculate several parameters derived from the fit: the minimum radius, 
$\rmin \equiv \mu F$; the weighted mean radius, \rmean, of the BLR clouds; and 
the virial factor, \fvir\ (Equation~\ref{eq:fvir}).

\begin{figure*}
\begin{center}
\includegraphics[width=0.7\textwidth]{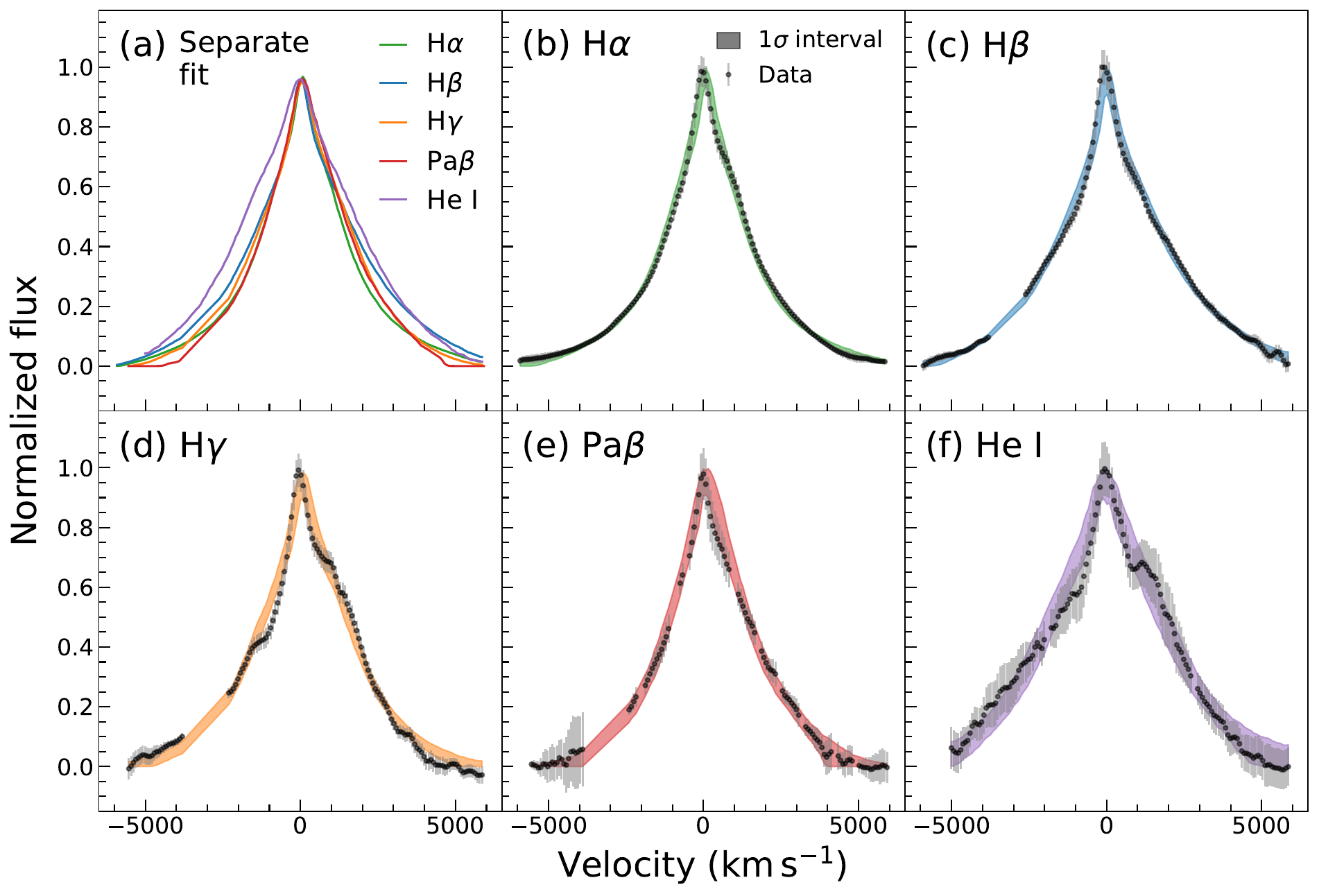}
\end{center}
\caption{The median and 68\% confidence intervals of the model line profiles 
when the lines are fitted separately.  The model profiles are overplotted in 
panel (a), and the 68\% confidence intervals of (b) \ha, (c) \hb, (d) \hg, 
(e) \pab, and (f) \hei\ are compared with the data (black).  The masked data are 
not plotted, so gaps are visible in some profiles.}
\label{fig:lsep}
\end{figure*}

\begin{figure*}
\centering
\includegraphics[width=0.85\textwidth]{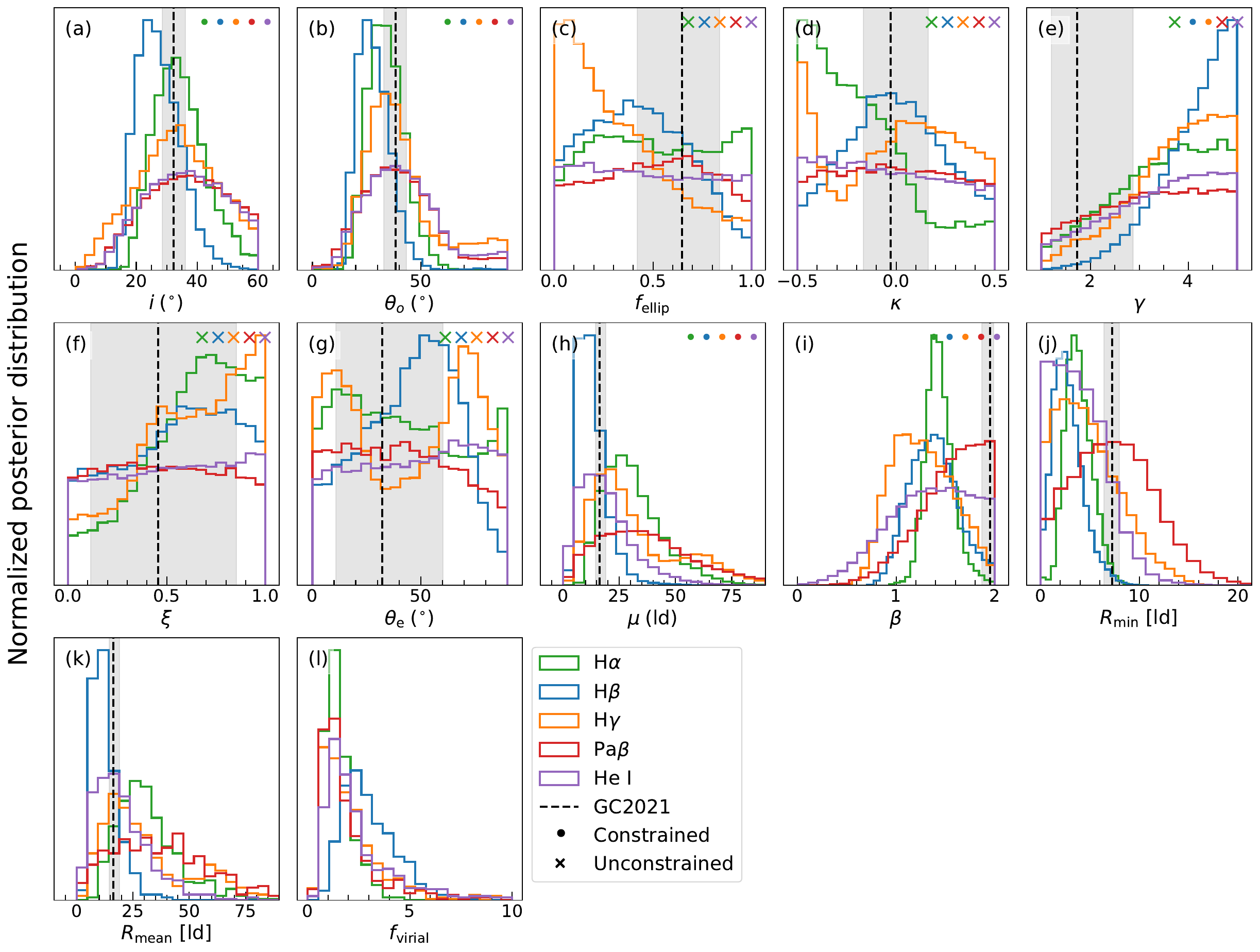}
\caption{Posterior probability distributions of the five lines fitted separately 
shown in colored histograms. For comparison, the dashed vertical line and 
the shaded region indicate the model inference best-fit results and 1-$\sigma$ 
intervals from the SARM joint analysis \citep{GC2021a}.  Panels (a)--(i) present 
the physical model parameters that are directly sampled in the fitting.  
The circles and crosses in these panels indicate whether these parameters are 
constrained by the data for each line profile (by comparing the width of their 
posterior distributions with their prior range).  The remaining panels present 
parameters that are derived from the posterior samples.  The y-axis tick labels 
are unimportant, so we remove them for clarity.  Details are discussed in 
Section~\ref{ssec:sep}.} 
\label{fig:psep}
\end{figure*}

\begin{table*}
\begin{center}
\caption{Summary of the combined fitting results}
\label{tab:fit_result}
\renewcommand{\arraystretch}{1.5}
\begin{tabularx}{.6\linewidth}{p{0.5cm} X | p{0.5cm} X | p{0.5cm} X | p{0.5cm} X}
\hline
\hline
\multicolumn{8}{c}{Tied Parameters} \\ \hline
$i\:(\degree)$        & $27.1_{-4.5}^{+4.5}$   & 
$\theta_o\:(\degree)$ & $28.8_{-4.4}^{+4.6}$   & 
$f_\mathrm{ellip}$    & $0.34_{-0.18}^{+0.22}$ & 
$\kappa$              & $0.03_{-0.15}^{+0.13}$ \\ 
$\gamma$              & $4.26_{-0.77}^{+0.51}$ & 
$\xi$                 & $0.86_{-0.22}^{+0.11}$ & 
$\theta_e$            & $31.9_{-18.8}^{+26.6}$ &          
                      &  \\ 
\hline
\end{tabularx}
\begin{tabularx}{.6\linewidth}{rXXXXX}
\hline
\multicolumn{6}{c}{Free parameters} \\ \cline{2-6}
                           & \mcc{\ha}              & \mcc{\hb}              & \mcc{\hg}              & \mcc{\pab}             & \mcc{\hei}             \\ \hline
$\mu$~(ld)                 & $20.6_{-5.7}^{+7.1}$   & $14.1_{-3.9}^{+5.0}$   & $17.2_{-4.7}^{+5.9}$   & $19.0_{-5.4}^{+7.1}$   & $10.3_{-2.9}^{+3.7}$   \\
$\beta$                    & $1.31_{-0.12}^{+0.13}$ & $1.44_{-0.25}^{+0.25}$ & $1.09_{-0.23}^{+0.29}$ & $1.42_{-0.39}^{+0.35}$ & $1.05_{-0.43}^{+0.50}$ \\
$F$                        & $0.15_{-0.04}^{+0.03}$ & $0.25_{-0.07}^{+0.06}$ & $0.20_{-0.12}^{+0.10}$ & $0.31_{-0.13}^{+0.12}$ & $0.38_{-0.23}^{+0.22}$ \\
$R_\mathrm{min}$~(ld)      & $3.1_{-1.1}^{+1.3}$    & $3.5_{-1.3}^{+1.6}$    & $3.3_{-2.0}^{+2.4}$    & $5.9_{-2.8}^{+3.2}$    & $3.8_{-2.4}^{+2.6}$    \\
$R_\mathrm{mean}$~(ld)     & $20.0_{-5.9}^{+8.2}$   & $13.6_{-4.0}^{+5.6}$   & $17.0_{-5.0}^{+6.6}$   & $18.4_{-5.2}^{+8.1}$   & $10.2_{-3.2}^{+3.6}$   \\
$f_\mathrm{virial}$        & $2.1_{-0.6}^{+0.9}$    & $2.5_{-0.7}^{+1.1}$    & $2.7_{-0.8}^{+1.2}$    & $2.8_{-0.8}^{+1.3}$    & $3.7_{-1.2}^{+1.7}$    \\
\hline
\end{tabularx}
\end{center}
{\small \textbf{Note.} 
The upper part of the table consists of the parameters tied for the five lines, 
and the lower part presents the model parameters ($\beta$, $F$, and $\mu$) that 
are sampled freely for individual lines, and the corresponding derived 
quantities (\rmin, \rmean, and \fvir).  The median and 68\% confidence 
interval values are reported.
}
\end{table*}

\subsection{Fitting the line profiles separately}
\label{ssec:sep}

In this section, we investigate what can be learned from fitting the five line 
profiles separately.  
We compare the data and the fitting results in Figure~\ref{fig:lsep}. 
We generate 500 model line profiles with parameters randomly selected from 
the posterior samples of each line, and plot the median line profiles in panel 
(a) and the 68\% confidence interval of each line in panels (b)--(f).  
Qualitatively, the BLR model can fit the line profiles reasonably well.  
The median model profiles reflect the differences of the lines.  The 68\% 
confidence profiles largely enclose the data of all lines.  The \ha\ and \hb\ 
model profiles are tighter constrained than the other three lines thanks to their 
smaller uncertainties.  The most obvious mismatch comes from the \hg\ line, due 
to the ``shoulder'' on the red wing as noted in Sec.~\ref{sssec:res}.  The \hei\ 
model profile shows some deviation from the data too, although always within 
the 1-$\sigma$ uncertainty level.  

The nonparametric parameters (line widths, A.I., and K.I.) for the model 
profiles are shown as vertical dotted lines in Figure~\ref{fig:npar}.  The model 
line widths follow the corresponding lines remarkably well in all cases.  
The A.I. values show clear differences to the data, although largely within 
the uncertainties.  In particular, the A.I. of \hg\ is much higher than 
the model value.  The K.I. values are consistent with the data, although 
the model posteriors tend to be higher.

There are 12 free parameters for each profile.  The posterior 
distributions are displayed in Figure~\ref{fig:psep}.  Following 
\cite{Raimundo2020}, we consider a model parameter constrained if its 
68~per~cent confidence range is less than half of its prior range.  Consistent 
with these authors, we find the geometric parameters, $i$, $\theta_o$, $\mu$, 
and $\beta$, of most of the lines can be constrained, while the remaining 
parameters, $f_\mathrm{ellip}$, $\kappa$, $\gamma$, $\xi$, and $\theta_e$, are 
mostly not.  The $i$ and $\theta_o$ of \ha\ and \hb\ agree well with 
the inclination derived by the SARM joint analysis  \citep{GC2021a}, although 
these two parameters are less well constrained for \hg, \pab, and \hei.  
In particular, the large asymmetry of \hg\ leads to a higher probability of 
a high inclination angle, so we see tentative double-peaked posterior 
distributions for $i$ and $\theta_o$.  In contrast, the distributions for \pab\ 
and \hei\ are broad and smooth, likely because the uncertainties of these two 
profiles are relatively large.  The fits have yielded generally small values for 
$\beta$ (except \pab), compared to the SARM result ($\beta \approx 1.95$).
This differs from the high value found in the SARM joint analysis but is 
consistent with the RM result for \hb\ reported by \citep{Bentz2021a}. 
Similarly, the minimum cloud radii (\rmin) for all lines except \pab\ prefer 
smaller values than the SARM result.  Nevertheless, the posteriors of 
the BLR sizes (\rmean) are largely consistent with the SARM results.

\subsection{Fitting the line profiles simultaneously}
\label{ssec:sim}

\begin{figure*}
\begin{center}
\includegraphics[width=0.7\textwidth]{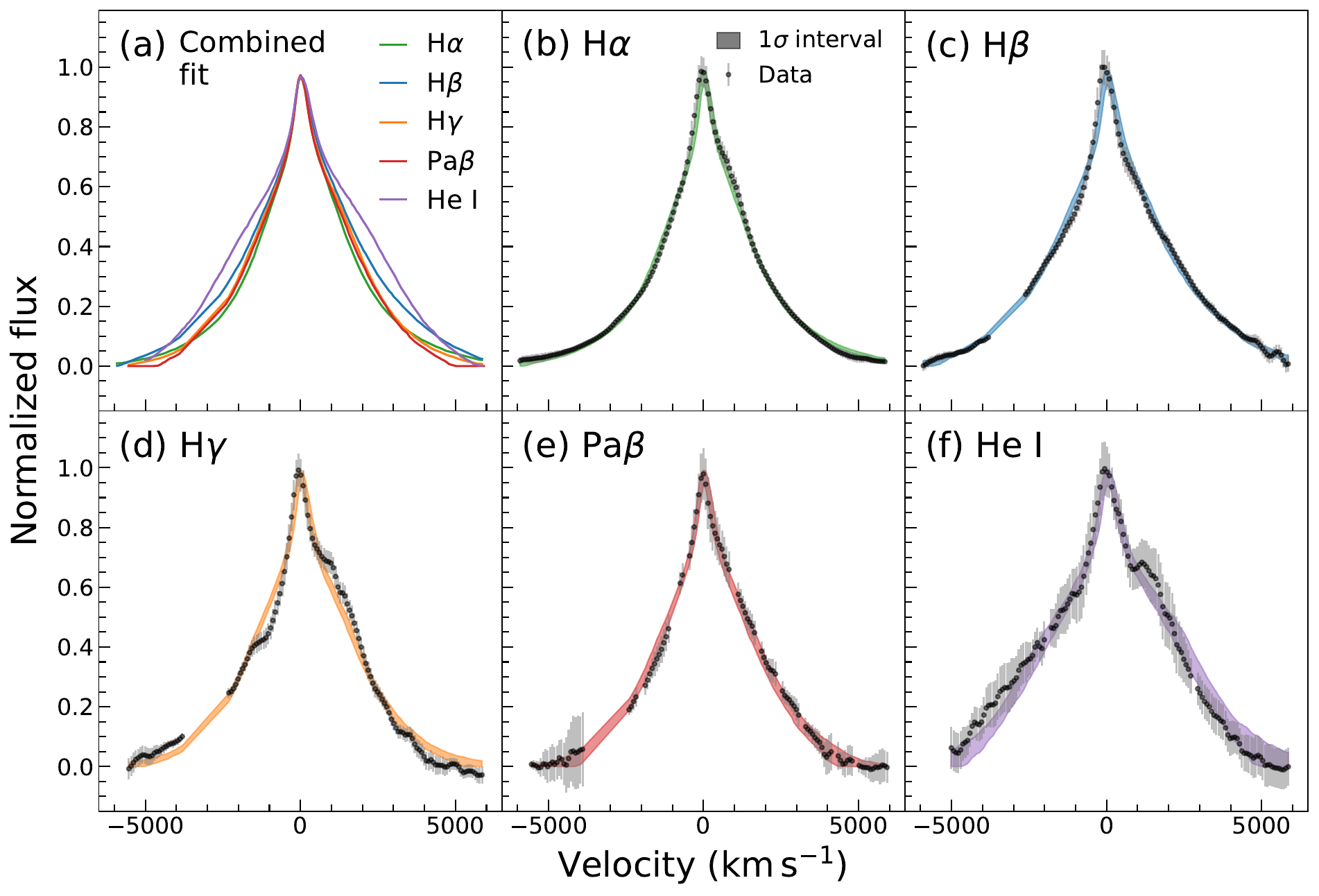}
\end{center}
\caption{The median and 68\% confidence intervals of the model line profiles 
when the lines are fitted simultaneously with most of the parameters tied.  
The panels and symbols are the same as that of Figure~\ref{fig:lsep}.}
\label{fig:ltie}
\end{figure*}

We perform a fit in which we tie all of the parameters of the BLR model for 
the five lines, except those defining the radial distribution of the clouds, 
namely, $\mu$, $\beta$, and $F$.  In this approach, we can test whether 
the difference in the line profiles can be solely explained by the radial 
stratification of the BLR.  There are 24 free parameters in the fit: 
9 tied between all the line profiles and 3 left separate for each of the 5 lines. 
We report the median and 68\% confidence interval values of the combined 
fit posterior samples in Table~\ref{tab:fit_result}.  The model profiles and 
the 68\% confidence intervals are shown in Figure~\ref{fig:ltie}. The panel (a) 
shows that the model profiles of \ha, \hg, and \pab\ are very similar, while 
\hb\ and \hei\ are wider.  This is similar to the results of the separate 
fitting.  Because most of the model parameters are tied now, we can conclude 
that the line profile differences can be explained by the radial distributions 
of the line emission.  Moreover, the combined fit results show tighter 68\% 
confidence intervals than the separate fit.  The improvement is the most obvious 
in \hg, \pab, and \hei\ lines whose uncertainties are relatively large.  
As a result, the deviation between the model and data of \hg\ is more obvious. 

The nonparametric values for these model profiles are similar to those from the 
separate fits, while the uncertainties of the simultaneous fits are smaller than 
those of the separate fits.  Again, they reproduce the line widths of the data 
well.  The A.I. values are similar between the lines because the model 
parameters controlling asymmetry in the model ($\kappa$ and $\xi$) are tied.
We verified that if these are left free, the model yields different values 
of $\xi$ to fit the individual asymmetries of the lines better.  Meanwhile, 
the other parameters do not change substantially, and in neither case is 
the high asymmetry of \hg\ reached.  The K.I. values of the tied model profiles 
are more consistent with the data than those of the separate fits.  
We note that both the model profiles yield consistent 
$\sigma_\mathrm{line}$ to the data profiles of all lines when fitting the data 
both separately and simultaneously.

\begin{figure*}
\centering
\includegraphics[width=0.85\textwidth]{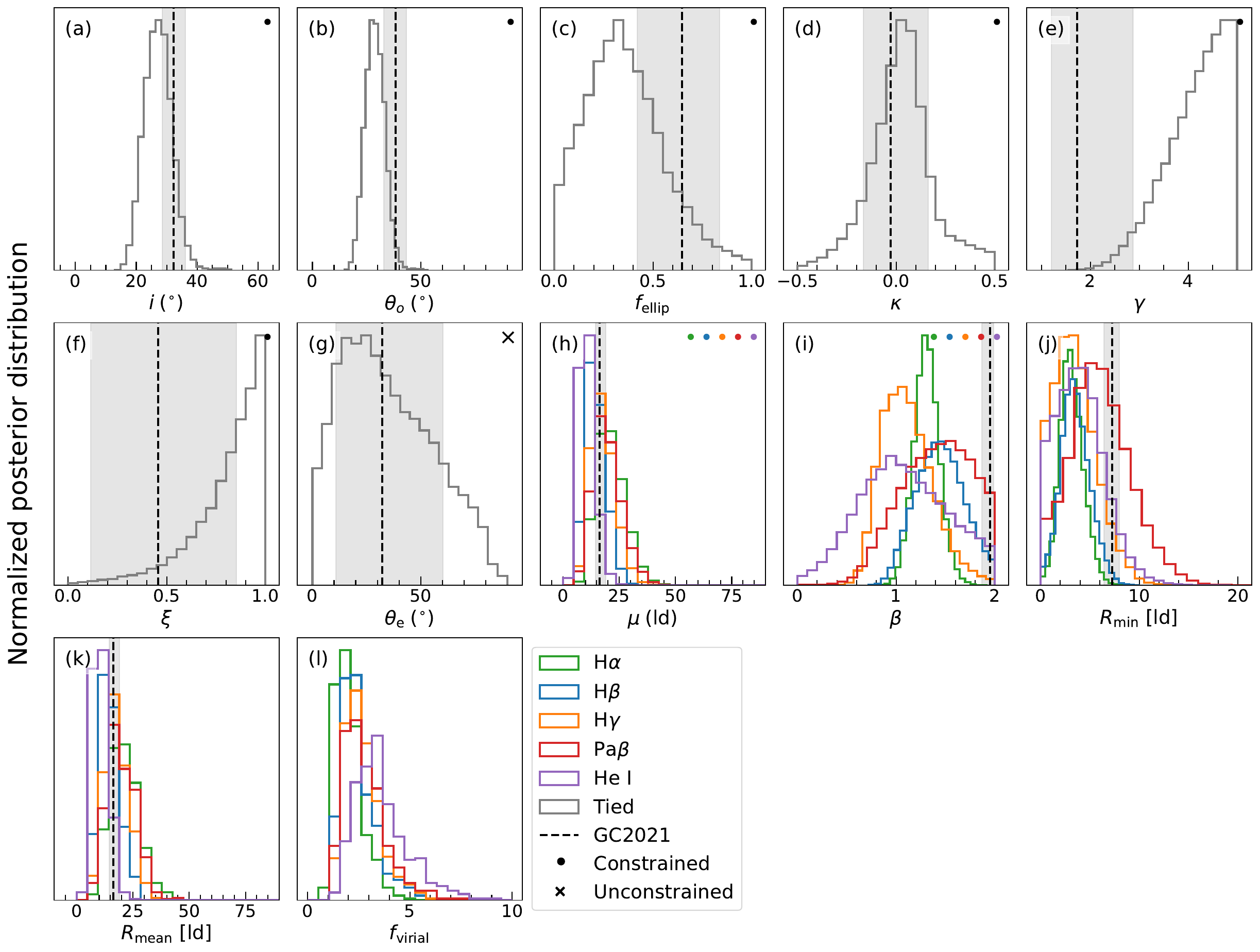}
\caption{Posterior probability distribution of the five lines fitted 
simultaneously with tied geometry parameters. The panels and symbols are similar 
as that of Figure~\ref{fig:psep}.  For comparison as before, the dashed vertical 
line and the shaded region indicate the model inference results from the joint 
SARM fit \citep{GC2021a}.}
\label{fig:ptie}
\end{figure*}

Figure~\ref{fig:ptie} shows that almost all of the tied model parameters are 
properly constrained, and consistent with the SARM joint analysis.  
In particular, the distributions of $\beta$ are similar to those from 
the separate fits, except that \pab\ now too prefers a lower value and so 
matches the other lines.  The lines also have similar \rmin$\sim4$~ld, hence, 
the line emission is concentrated in an inner ring and extends to large radii.
In passing, we tested to further tie \rmin\ in the fit and found 
the results stay almost entirely the same with the tied 
$\rmin\approx 3.5~\mathrm{ld}$, confirming that the simultaneous fit favors 
the different lines sharing the same \rmin.  \cite{GC2021a} tested fixing 
the $\rmin=4~\mathrm{ld}$ in the SARM analysis and found a reasonable fitting 
result with $\beta \approx 1$, interestingly, close to our $\beta$.  However, 
\cite{GC2021a} caution that additional restrictions bias their geometric 
distance measurement by about 30\%.

The parameter that differs most from the SARM result is $\gamma$, where a large 
value is preferred, indicating that the vertical distribution of line emission 
is more concentrated towards the surface of the BLR.  However, there is 
a long tail towards low values in the posterior distribution of $\gamma$.  
The number of clouds in the mid-plane increases by less than $20\%$ when 
$\gamma$ decreases to 1.7 (the value derived in the SARM joint analysis), so 
the model does not change substantially.  Nevertheless, the edge-concentrated 
structure with high $\gamma$ may indicate that a biconical wind-like structure 
\citep[e.g.][]{Matthews2016,Waters2021} could fit the line profiles.  
The flexibility enabled by including such a capability in the model may enable 
it to better reproduce the asymmetric features in broad lines of NGC~3783.

\section{BLR geometry and virial factor from single-epoch line profiles}
\label{sec:se}

\subsection{BLR model parameters}
\label{ssec:phys}

In this section, we discuss the model parameters of the BLR that can be 
measured from single-epoch line profiles.  For individual line profiles, we can 
constrain the inclination ($i$) and disk thickness ($\theta_o$), a finding 
consistent with \cite{Raimundo2019,Raimundo2020}.  Because we fix the BH mass, 
the radial distribution of the line emission can be constrained too.  More 
interestingly, we find the BLR model can simultaneously fit multiple line 
profiles using the same geometry and kinematics.  Almost all of the model 
parameters can be constrained, and only the radial distributions change for each 
line, indicating that the radial emissivity distributions are somewhat 
different, as expected (see below).  The \HI\ lines tend to favor heavy-tailed 
distributions with $\beta>1$, while the \hei\ line is close to exponential 
($\beta \approx 1$).  While it is beyond the scope of this work, we note 
that $\beta \gtrsim 1$ is close to a truncated power-law distribution that 
the photoionization model may produce \citep[e.g.][]{Netzer2020}.

\begin{figure*}
\centering
\includegraphics[width=0.75\textwidth]{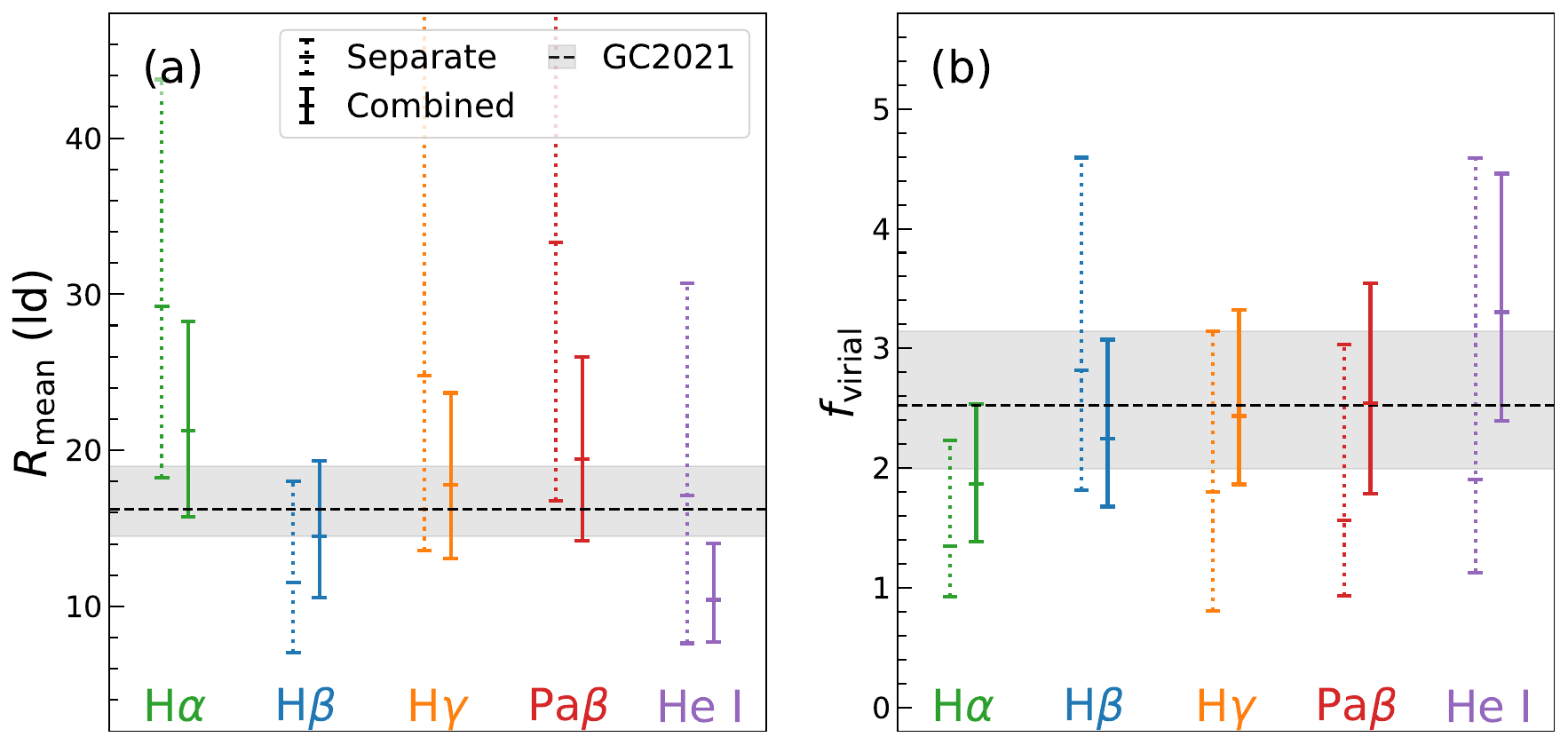}
\caption{Comparison of derived parameters. 
Values of (a) mean radius and (b) virial factor are shown for 
the separate (dotted lines) and combined (solid lines) fits.  The three 
horizontal bars on each line correspond to the 16th, 50th, and 84th percentiles 
of the posterior distribution.  The dashed lines and grey regions represent 
the results from the SARM joint analysis \citep{GC2021a}.}
\label{fig:phys}
\end{figure*}

We now compare the BLR radius (\rmean) of the combined fit to the five 
lines (Figure~\ref{fig:phys}a) with published measurements based on the \hb\ and 
\brg\ lines \citep{Bentz2021a,Bentz2021b,GC2021a,GC2021b}.  We focus on 
the joint SARM analysis because it obtains the tightest constraints of the model 
parameters using the spectro-astrometry and RM data simultaneously, and also 
because we have adopted its BH mass here.  The \rmean\ from the separate 
fitting show relatively large uncertainties.  The \rmean\ of \hb, which happens 
to have the smallest uncertainty, is consistent with the \rmean\ derived by 
the SARM analysis.  In contrast, \rmean\ derived from the combined fitting show 
much smaller uncertainties, with the \hb\ \rmean\ consistent with the SARM 
result as well.  The \hei\ \rmean\ is the smallest among the five lines.  
We note that the posterior distributions of \rmean\ are strongly correlated 
between different lines (Figure~\ref{fig:cplot_sim}).  While such a correlation 
increases the uncertainties of the mean radii ($\gtrsim 30\%$, 
Figure~\ref{fig:phys}a), as discussed below, it also means that the ratios of 
the mean radii have smaller uncertainties ($\lesssim 10\%$).

As shown in Table~\ref{tab:line_ratio}, we measure the BLR mean radius 
ratios of \ha:\hb:\hg:\pab:\hei\ from the simultaneous fitting to be 
1.47:1.0:1.22:1.36:0.72.  They are largely consistent with the RM observation 
results reported by \cite{Bentz2010b}.  Our relative lags of \hg\ and \hei\ 
lines are larger than the RM results.  We also calculate the time lag ratios 
based on the radiation pressure confined (RPC) BLR model as described by 
\cite{Netzer2020}.  
The theoretical model calculation explores a range of parameters that are in 
agreement with most RM measurements of various hydrogen and helium lines 
\citep[e.g.][]{Bentz2009,Bentz2010b}.  The most important parameters of 
the RPC model are the radial dependence of the covering factor of the clouds, 
\rmin\ (which is somewhat arbitrary), the BLR outer radius, which is determined 
by graphite dust sublimation radius, gas metallicity, and turbulent velocity 
within individual clouds.  The level of ionization, and hence line emissivity at 
all locations, are obtained naturally from the assumption that the clouds' 
column densities are large, and they are in total gas and radiation pressure 
equilibrium.  The mean emissivity radius for each line is then computed from 
the model and then translated to time lags, which depend on the light curve of 
the driving continuum.  All such models predict 
$\tau_\mathrm{H\alpha}>\tau_\mathrm{H\beta}>\tau_\mathrm{H\gamma}>\tau_\mathrm{He~I}$ 
and the range is illustrated in Table~\ref{tab:line_ratio}.  Similar tendencies 
are also predicted by the very different LOC model computed by \cite{Korista2004}.

We suspect our large \hg\ radius is due to the error of the \hg\ line profile. 
The \hg\ line, the weakest among the three Balmer lines, is blended with \OIIId\ 
line (Figure~\ref{fig:spec}a).  Therefore, the small but systematic 
residual of the decomposition may lead to a too large \hg\ BLR size compared 
with its actual dimensions.  The clear systematic deviations between the data 
and model in Figure~\ref{fig:ltie}d support this point to some extent.  This 
issue highlights the importance of high-quality line profiles to reveal robust 
BLR properties.  Another effect that may influence the observed line emission 
distribution is the polar dust around the BLR \citep{Honig2013}.  Higher 
extinction in the center will make the BLR look more extended.  Since extinction 
is larger at shorter wavelengths \citep{Li2007}, this effect might enlarge the 
observed \hg\ size more than \ha\ and \hb.  It is worth mentioning that 
\cite{Bentz2021a} 
reported a very small time lag of 3.7~ld for \hg\ in NGC~3783, as small as that 
of He~{\sevenrm II}.  As discussed by the authors, the \hg\ time lag was likely 
underestimated due to imperfect internal flux calibration using \OIIIb\ line 
flux.  Observations of more targets would be useful to disentangle these effects 
on the measured BLR size across different lines.  

We remind the readers that the models shown here do not include 
time-dependent variations of the ionizing source and we caution that the flux 
weighted radius (\rmean) may be different from the time lag depending on 
the structure of the BLR and the ionization continuum light curve 
\citep{Netzer1990}.  However, we expect the \rmean\ ratios from our simultaneous 
fitting to be close to the ratios of the time lags because the different lines 
are assumed to share the same BLR structure.
To conclude, our analysis shows that different broad emission lines appear to 
share most of the geometry and kinematics of the BLR.  The relative BLR sizes 
are largely consistent with the theoretical expectation of photoionization models.  
This suggests that one can combine spectro-astrometry and RM data of different 
lines in a joint analysis by sharing most of the BLR model parameters of the two 
lines.

\begin{table}[]
\begin{center}
\caption{BLR size ratios ratios}
\label{tab:line_ratio}
\renewcommand{\arraystretch}{1.5}
\begin{tabular}{cccc}
\hline\hline
Line  & This work                            & RM observation & RPC model \\
(1)    & (2)                                 & (3)            & (4)       \\ \hline
\ha    & $1.46_{-0.12}^{+0.12}$ & 1.54           & 1.4--1.5  \\
\hb    & 1.0                    & 1.0            & 1.0       \\
\hg    & $1.22_{-0.10}^{+0.11}$ & 0.61           & 0.8-0.9   \\
\pab   & $1.35_{-0.15}^{+0.15}$ & --             & 1.3-1.4   \\
\hei   & $0.72_{-0.11}^{+0.11}$ & 0.36           & 0.6-0.8   \\
\hline
\end{tabular}
\end{center}
{\small \textbf{Note.} 
Column (1): Line names.
Column (2): The BLR size ratios derived from our simultaneous fitting.  
We randomly selected 500 sets of model parameters from the posterior samples and 
calculated the median \rmean\ ratios and the 68\% confidence intervals.
Column (3): The time lag ratios measured by \cite{Bentz2010b}.  The \pab\ time 
lag is not measured.
Column (4): The time lag ratios derived by the radiation pressure confined (RPC) 
BLR model as described by \cite{Netzer2020}.  We explored a range of model 
parameters that are in agreement with most RM measurements of various hydrogen 
and helium lines \citep[e.g.][]{Bentz2009,Bentz2010b}.
The \hb\ time lag is normalized to unity in all of the results.
}
\end{table}

\subsection{The virial factor}
\label{ssec:fvir}

The virial factor encapsulates the geometry of the BLR by linking the BH mass to 
the measurable properties of BLR size and line width. In this paper, we define 
it as
\begin{equation}
\label{eq:fvir}
f_\mathrm{virial} = \frac{G M_\mathrm{BH}}{\sigl^2 \rmean},
\end{equation}
where \sigl\ is the second moment of the model line profile, \rmean\ is 
the mean radius of the BLR, and \mbh\ is the fixed BH mass.  A key property is 
that \fvir\ scales with $\mbh/r$, where $r$ indicates the BLR size.
Although we have fixed the BH mass of NGC~3783, the value of $f_\mathrm{virial}$ 
that we derive does not depend on \mbh\ because the BLR radius is a free 
parameter in the fit.  The reason, explained in Section~\ref{ssec:dybel}, 
is that \mbh\ and $\mu$ (or \rmean) are fully degenerate in our model: these 
parameters scale together without changing the line profile.  Therefore, we can 
expect to derive meaningful virial factors from the fit.  To confirm this point, 
we fit the data with a fixed $\mbh=10^{6.4}\,\Msun$ (10 times smaller) and got 
the same \fvir\ results.  

As shown in Figure~\ref{fig:phys}b, the derived values from both 
separate fit and combined fit are consistent with that of the SARM joint 
analysis ($\fvir=2.52_{-0.53}^{+0.62}$).  The combined fit shows 
smaller uncertainties than the separate fit.  This is an encouraging result for 
investigating the BLR dynamics and measuring the BH mass.  Our method could 
enable one to constrain the individual virial factor for each AGN by modeling 
the broad emission line(s) without the need for dynamical modeling of RM data 
\citep[e.g.][]{Villafana2023}.  As a practical approach, one can fix the BH mass 
according to the single-epoch estimate (based on an averaged virial factor) 
\citep{McLure2001,Ho2015,DallaBonta2020} and model the broad line profile(s) to 
derive the \fvir\ for individual AGNs.  The \fvir\ can then be used to refine 
the BH mass estimate.  As such, it could reduce the uncertainty in the BH mass 
that is otherwise introduced by adopting an average virial factor 
\citep{Collin2006,Shen2014}.  We caution that the derived virial factor 
may be biased by the oversimplified BLR model, the influence of which will be 
investigated with many more sources in the future.

\subsection{Caveats}
\label{ssec:cav}

In this work, we investigate the BLR structure by modeling 
the normalized single-epoch line profiles.  We model the broad-line emission and 
the associated kinematics with a Monte Carlo model of points without a physical 
size.  This model is intended to avoid considering the details of the photoionization 
physics and cannot predict the line strength physically according to the AGN 
luminosity.  Without the data spatially resolving the BLR structure (e.g. 
GRAVITY differential phase), the model parameters may be degenerate when only 
fitted with the line profiles.  Interestingly, for NGC~3783, we find 
the single-epoch line profiles, especially fitted simultaneously, can provide 
most of the BLR model parameters consistently with the fitting including 
the size measurements.  We caution, however, that more studies on different BLRs 
are needed to understand whether the conclusion holds widely.

To test whether our BLR model is quantitatively plausible with 
photoionization physics, we estimate the \hb\ luminosity assuming the Case~B 
recombination \citep{Osterbrock2006} and the geometric covering factor based on 
our model fitting result.  With $\theta_o \approx 28.8^\circ$ 
(Table~\ref{tab:fit_result}), the geometric covering factor is 
$\sin \theta_o \approx 0.5$.  We adopt the ratio of \hb\ line and hydrogen 
recombination coefficients of 
$\alpha^\mathrm{eff}_\mathrm{H\beta}/\alpha_B \approx 1/8.5$ and the UV photon flux 
$\int_{\nu_0}^\infty\frac{L_\nu}{h\nu}d\nu \approx$~1.6--3.7~$\times 10^{53}\,\mathrm{s}^{-1}$  
($\nu_0=13.6\,\mathrm{eV}$), which is estimated with the measured 
$\lambda L_\lambda(5100\,\mathrm{\AA}) \approx 4.1 \times 10^{42}\,\mathrm{erg\,s^{-1}}$ 
and the assumed AGN SED with low and intermediate Eddington ratios 
\citep{Ferland2020,Jin2012} to enclose the range of typical Seyfert galaxy SEDs. 
We derive $L_\mathrm{H\beta}\approx$~0.4--0.9~$\times 10^{41}\,\mathrm{erg\,s^{-1}}$. 
The measured \hb\ luminosity\footnote{The measured 
$\lambda L_\lambda(5100\,\mathrm{\AA})$ and $L_\mathrm{H\beta}$ are both from 
the X-shooter spectrum used in this work, so they may share the same systematic 
flux uncertainty, which does not influence our comparison of \hb\ luminosity.}, 
$\sim 1.1 \times 10^{41}\,\mathrm{erg\,s^{-1}}$, 
is comparable to the estimated range, if not slightly higher, indicating that 
our BLR model is plausible in terms of line luminosity. While this estimate is 
admittedly oversimplified, it illustrates that the shape of the SED may easily 
influence the line luminosity by a factor of a few.  It is worth noting that we 
assume the maximum absorption of the UV photons with the model BLR geometry, and 
Case~B recombination does not consider the self-absorption of the \hb\ photons.  
More detailed photoionization calculations with CLOUDY may only provide weaker 
line emission, which reflects the aforementioned problem in 
Section~\ref{ssec:dybel}.  Although this problem is beyond the scope of this 
work, our method provides a new approach to address it with single-epoch spectra.


\section{Conclusions}
\label{sec:con}

We investigate the BLR structure of NGC~3783 using multiple broad lines in 
a high-resolution single-epoch spectrum obtained with VLT/X-Shooter.  
We decompose the strongest five broad lines (\ha, \hb, \hg, \pab, and 
\hei~$\lambda$5876), and model their profiles using the newly developed tool 
\dybel, which allows one to tie parameters of the dynamical model between the 
lines.  Since the BH mass and the BLR radius are fully degenerate, we opt to fix 
the BH mass to a value reported in the literature and focus on the BLR structure 
and emissivity that can be derived from the line profiles.  In the future, more 
comprehensive analyses will be useful to explore the potential of this method, 
with more broad-line AGNs, different spectral resolutions and signal-to-noise 
ratios, and different BLR dynamical models.  Our main results are, 

\begin{enumerate}
\item All lines analyzed here show broader wings than a Gaussian profile 
and are asymmetric (skewed to the red side). The \hei~$\lambda$5876 profile is 
broader than the hydrogen profiles studied here.

\item We develop a fitting tool to model the line profiles with a dynamical BLR 
mode. Fitting multiple lines simultaneously by tying many of their parameters 
together yields a solution that is better constrained than when fitting them 
individually.  In particular, it yields useful constraints on some parameters 
such as inclination, BLR size, and the virial factor. 

\item The difference in line profiles can be explained almost entirely in terms 
of differing radial distributions of the line emission.  The derived relative 
BLR time lags are mostly consistent with the RM observation and with theoretical 
model calculations.  Our results support that it is possible to combine 
spectro-astrometry and RM data in a joint analysis.

\item The virial factor we derive is nearly the same for the five lines and is 
independent of the adopted BH mass.  We argue that by enabling one to constrain 
the virial factor for an individual AGN using a single epoch spectrum, this 
method can reduce the uncertainty in BH masses derived from single epoch spectra. 

\end{enumerate}

\begin{acknowledgements}
L.K. thanks the financial support from the Mitacs Globalink Research Award. 
AWSM and LK acknowledge the support of the Natural Sciences and Engineering 
Research Council of Canada (NSERC) through grant reference number 
RGPIN-2021-03046. 
J.S. dedicates this paper in memory of his grandmother, G. X. (1930-2023).
J.S. thanks Leonard Burtscher for sharing the X-Shooter data.
\end{acknowledgements}

%
%

\bibliography{ref}{}
\bibliographystyle{aa}

\begin{appendix}

\section{Simultaneous fitting results with all parameters tied}

\begin{figure*}
\centering
\includegraphics[width=0.9\textwidth]{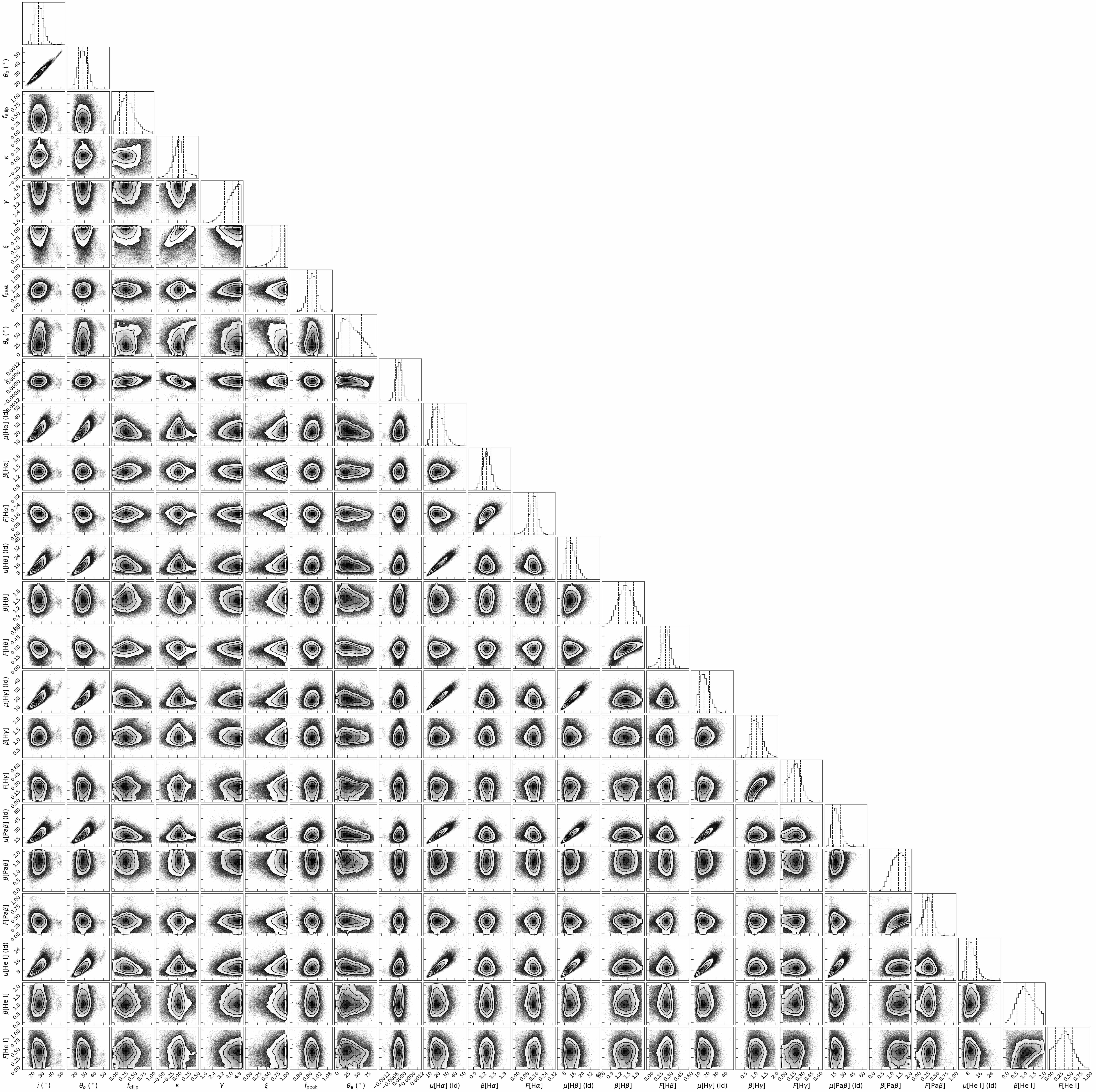}
\caption{Corner plot of the simultaneous fitting.  The first nine parameters are 
tied in the fitting, while the remaining parameters are fitted for individual 
lines.}
\label{fig:cplot_sim}
\end{figure*}

The corner plot of the simultaneous fitting is shown in 
Figure~\ref{fig:cplot_sim}.  The separate fittings show similar but less 
constrained results.  We opt not to show all of the corner plots of the separate 
fitting for simplicity because all of the useful information has been shown in 
Figure~\ref{fig:psep}.

\end{appendix}

\end{document}